\documentclass[10pt, twocolumn]{IEEEtran}

\usepackage{amsmath}
\usepackage{amsfonts}
\usepackage{color}
\usepackage{graphicx}
\usepackage{mathrsfs}
\usepackage{times}
\usepackage{empheq}
\usepackage{ifpdf}
\ifpdf
\usepackage{epstopdf}
\fi
\usepackage{cite}
\usepackage{relsize}
\usepackage{arydshln}
\usepackage[normalem]{ulem}
\usepackage{amssymb}
\usepackage{subcaption}
\usepackage{enumerate}
\title{Spectrum Efficient MIMO-FBMC System using Filter Output Truncation}
\author{Adnan Zafar, Lei Zhang, Pei Xiao and Muhammad Ali Imran
\thanks{A. Zafar and P. Xiao are with Institute for Communication Systems (ICS), University of Surrey, Guildford, UK. Emails: \{a.zafar, p.xiao\}@surrey.ac.uk. A. Zafar is also affiliated with Institute of Space Technology, Islamabad, Pakistan. Email: adnan.zafar@ist.edu.pk}
\thanks{L. Zhang and M. A. Imran are with School of Engineering, University of Glasgow, Glasgow, UK. Email: \{lei.zhang, muhammad.imran\}@glasgow.ac.uk}
}

\makeatletter
\let\old@ps@headings\ps@headings
\let\old@ps@IEEEtitlepagestyle\ps@IEEEtitlepagestyle
\def\confheader#1{%
	\def\ps@headings{%
		\old@ps@headings%
		\def\@oddhead{\strut\hfill#1\hfill\strut}%
		\def\@evenhead{\strut\hfill#1\hfill\strut}%
	}%
	\def\ps@IEEEtitlepagestyle{%
		\old@ps@IEEEtitlepagestyle%
		\def\@oddhead{\strut\hfill#1\hfill\strut}%
		\def\@evenhead{\strut\hfill#1\hfill\strut}%
	}%
	\ps@headings%
}
\makeatother

\confheader{%
	This article has been published in IEEE journal. The final version is available at http://dx.doi.org/10.1109/TVT.2017.2771531
}
\begin{document}
\maketitle
\begin{abstract}
Due to the use of an appropriately designed pulse shaping prototype filter, filter bank multicarrier (FBMC) system can achieve low out of band (OoB) emissions and is also robust to the channel and synchronization errors. However, it comes at a cost of long filter tails which may reduce the spectral efficiency significantly when the block size is small. Filter output truncation (FOT) can reduce the overhead by discarding the filter tails but may also significantly destroy the orthogonality of FBMC system, by introducing inter carrier interference (ICI) and inter symbol interference (ISI) terms in the received signal. As a result, the signal to interference ratio (SIR) is degraded. In addition, the presence of intrinsic interference terms in FBMC also proves to be an obstacle in combining multiple input multiple output (MIMO) with FBMC. In this paper, we present a theoretical analysis on the effect of FOT in an MIMO-FBMC system. First, we derive the matrix model of MIMO-FBMC system which is subsequently used to analyze the impact of finite filter length and FOT on the system performance. The analysis reveals that FOT can avoid the overhead in time domain but also introduces extra interference in the received symbols. To combat the interference terms, we then propose a compensation algorithm that considers odd and even overlapping factors as two separate cases, where the signals are interfered by the truncation in different ways. The general form of the compensation algorithm can compensate all the symbols in a MIMO-FBMC block and can improve the SIR values of each symbol for better detection at the receiver. It is also shown that the proposed algorithm requires no overhead and can still achieve a comparable BER performance to the case with no filter truncation.
\end{abstract}
\vspace*{-0.3cm}
\section{Introduction}
\IEEEPARstart{F}{ilter} bank multicarrier (FBMC) has illustrated profound advantages over conventional multicarrier modulation (MCM) schemes such as orthogonal frequency division multiplexing (OFDM) in time and frequency dispersive channels \cite{5753092,6849669,6962232,4426233}. Such advantages come from the fact that OFDM suffers from large out of band (OoB) emissions and thus require large guard bands to protect neighboring channels, hence reducing the efficiency of the system. This presents a major source of problem that limits the applicability of OFDM in some present and future communication systems \cite{6568951}. FBMC, on the other hand, is a promising technique that overcomes this problem by utilizing a specially designed prototype filter such as isotropic orthogonal transform algorithm (IOTA) which is well localized both in time and frequency \cite{4449830}. This prototype filter enables FBMC to provide best OoB emission among the new waveforms proposed for future networks, such as generalized frequency division multiplexing (GFDM) \cite{5073571}, universal filtered multi-carrier (UFMC) \cite{6824990, 7829438}, filtered orthogonal frequency division multiplexing (FOFDM) \cite{7417854} and their variants \cite{staheri}. This advantage enables FBMC systems to utilize the fragmented spectrum more efficiently \cite{7390479}.\vspace*{.1cm}\\
\indent Other main advantage of FBMC include higher spectral efficiency compared to conventional OFDM systems. It is due to the good time and frequency localization properties of the prototype filter in FBMC that ensures inter carrier interference (ICI) and inter symbol interference (ISI) are negligible without the use of cyclic prefix (CP) \cite{5753092}. The strict synchronization requirements in conventional OFDM based systems are also much relaxed for FBMC system. This facilitates low complexity implementation of multi-user (MU) access in uplink transmissions for FBMC systems \cite{6629693,Mattera2015}.  Due to these advantages, FBMC is considered as a key area of research for the past several years and one of the most promising waveform candidate for future wireless networks \cite{6877912,6736749}.\vspace*{.1cm}\\
\indent Unlike conventional OFDM, the FBMC system utilizes orthogonal QAM symbols as the system is non-orthogonal in complex plane. However, FBMC requires more complex receiver structure, particularly when combined with MIMO as compared to the MIMO-OFDM based systems. Moreover, FBMC system may encounter residual interference terms in the form of ICI and ISI if a low complexity channel equalization is used for highly dispersive channels. The impact of doubly dispersive channel on a SISO-FBMC system with both zero forcing (ZF) and minimum mean squared error (MMSE) based one tap equalization schemes is analyzed in \cite{7549072}. It is proposed that a complex multi-tap equalization may be required as the performance of the FBMC system is severely limited by strong doubly dispersive channel impact. The authors in \cite{7870574} have investigated the performance degeneration of OFDM and FBMC systems in doubly-selective channels using a closed-form bit error probability (BEP) expression. It is shown that FBMC performs better than CP-OFDM in highly time-varying channels due to the use of well localized prototype filter.\vspace*{.1cm}\\
\indent Unlike OFDM, the use of FBMC in multi-antenna configurations is not as straightforward and the applications are very limited. Tensubam \textit{et al.} in \cite{tensubam2014study} have presented a study on recent advancements in MIMO-FBMC and suggest that filtered multitone (FMT) based FBMC systems offer the same flexibility as OFDM in adopting MIMO technology. However, it is spectrally inefficient compared to other variants of FBMC like cosine modulated multitone (CMT) and staggered modulated multitone (SMT) as it requires guard bands between the subcarriers. But unlike conventional OFDM system, the received symbols in CMT and SMT based FBMC systems are contaminated with pure imaginary intrinsic interference. This interference proves to be a huge obstacle in combining MIMO techniques with FBMC.\vspace*{.1cm}\\
\indent A two step receiver for MIMO-FBMC is proposed in \cite{6629839}, where the first stage estimates and cancels this intrinsic interference, while the second stage improves the estimation using widely linear processing. The authors in \cite{6362751} have shown strong correlation between the real and imaginary components of FBMC signal and have proposed a new equalizer structure by exploiting the imaginary intrinsic interference components. A scheme referred as FFT-FBMC, is proposed by Rostom Zakaria \textit{et al.} in \cite{6144767} and is applied to multiple antenna systems. Although, FFT-FBMC technique can address the issue of intrinsic interference by using a CP, however, it has a poor bit error rate (BER) performance as compared to the conventional OFDM systems. It was shown in \cite{7511672} that the FFT precoded signals in FFT-FBMC can reduce the frequency band occupied by each subcarrier by reducing the interference power in the immediate adjacent sub band as compared to conventional FBMC. Jayasinghe \textit{et al.} in \cite{6655423} have analyzed the effect of intrinsic interference in FBMC system and proposed a precoder based on signal to leakage plus noise ratio (SLNR) at the transmitter side to overcome its effects on the FBMC system. It is shown that the proposed precoder design at the transmitter outperforms the equalization based FBMC and OFDM systems. Recent developments in combining FBMC with massive MIMO are discussed in \cite{6965986}.\vspace*{.1cm}\\
\indent There has been investigations on the performance of MIMO-FBMC system in frequency selective channels. Various precoding and equalization techniques are proposed to achieve robustness against channel frequency selectivity and to improve the spectral efficiency in a MIMO-FBMC system \cite{6295673}. The authors in \cite{7707463} have presented a single-tap precoder and decoder design for multiuser MIMO-FBMC system for frequency selective channels by optimizing the MSE formula under ZF and MMSE design criteria. Mestre \textit{et al.} in \cite{7305795} have proposed a novel architecture for MIMO-FBMC system by exploiting the structure of the analysis and synthesis filter banks using approximation of an ideal frequency-selective precoder and linear receiver. Another precoding and decoding technique for MIMO-FBMC system is proposed in \cite{6638621} to enable multi-user transmission in frequency selective channels. Soysa \textit{et al.} in \cite{6843159} have evaluated the performance of precoding and receiver processing techniques for multiple access MIMO-FBMC system for an extended ICI/ISI scenario in uplink and downlink. A. Ikhlef \textit{et al.} in \cite{5291308} proposed successive interference cancellation (SIC) to extract the transmitted information in a MIMO-FBMC system. The proposed solution outperforms the classical one tap equalizers in case of moderate and high frequency selective channels. Chang \textit{et al.} in \cite{chang2010} have presented a precoded SISO-FBMC system without CSI at the transmitter. The proposed system is limited by the assumption of perfect equalization at the receiver whereas, imperfect equalization can lead to residual ISI and ICI terms. The authors then analyzed the effect of interference from imperfect equalization in \cite{6388232}. The results suggest that multi-tap equalization is required to reduce the effect of interference in FBMC system for highly frequency selective channels. Inaki Estella \textit{et al.} in \cite{5722365} provided a comparison between multi-stream MIMO based OFDM and FBMC systems and suggested that OFDM achieves a lower energy-efficiency than the FBMC. However, unlike OFDM, the use of multiple streams increases interference in FBMC which require new equalization techniques. Ana I. Perez-Neira \textit{et al.} have presented a detailed and comprehensive overview of various challenges in MIMO-FBMC systems and their known solutions in \cite{7491375}.\vspace*{.1cm}\\ 
\indent The aforementioned studies are mainly focused on the performance of MIMO-FBMC systems in frequency selective channels and its spectral efficiency compared to OFDM based systems. However, despite the fact that FBMC does not require a CP, it is not completely free from overhead as the filter bank itself introduces extra tails in the FBMC block that affects the spectral efficiency of the system. A recent study has considered improving the spectral efficiency in FBMC system by tackling the over head (tails) caused by the filter operation. The authors in \cite{7900335} have introduced non data symbols (virtual symbols) before and after each FBMC data packet for shortening the ramp-up and ramp-down periods. In \cite{7390479}, it is pointed out that filter output truncation (FOT) or tail cutting can improve the spectral efficiency of FBMC system but require one extra tail to be transmitted as overhead along with the FBMC block.\vspace*{.1cm}\\
\indent In this paper, we provided an in-depth analysis of FOT in a MIMO-FBMC system. We investigated the possibilities of completely discarding all the tails (overhead) to improve the spectral efficiency of the MIMO-FBMC system. To achieve this, we represented the complete MIMO-FBMC system in a matrix form including the filter operation, tail cutting/truncating, channel convolution, equalization, detection etc. The interference terms like ICI and ISI introduced by the FOT along with the intrinsic interference terms are then derived using the MIMO-FBMC matrix model. In light of the analytical results, we proposed a compensation algorithm to overcome the interferences caused by FOT. The proposed algorithm enables the complete elimination of the overhead in a MIMO-FBMC system by compensating the truncation affect at the receiver. As a result, the spectral efficiency of MIMO-FBMC systems is improved. The contributions of this paper are summarized as follows.
\begin{itemize}
\item We first derive a compact matrix model of MIMO-FBMC system which lays the ground for the subsequent in-depth analysis of the effect of FOT on the detection performance in terms of the SIR and BER.
\item Based on the matrix model, we then analyze the impact of finite filter length and different types of FOT on the system performance. Through simulation results, it is shown numerically that FOT can overcome the high overhead but significantly degrade the SIR of the symbols at the edges.
\item Thirdly, based on the observations made in the aforementioned numerical analysis, a compensation algorithm is designed to compensate the symbols in a MIMO-FBMC block to improve the SIR of each symbol. The advantage of the algorithm is that it requires no overhead but can still achieve a similar performance compared to the case with no FOT.
\end{itemize}

\textit{Notations:} Vectors and matrices are denoted by lowercase and uppercase bold letters. $\{.\}^H$ and $\{.\}^T$ represent conjugate transpose (hermitian conjugate) and transpose operations. $\mathcal{F}$ and ${\mathcal{F}}^H$ denote the normalized N point DFT and IDFT matrices. $A\otimes B$ represents kronecker product of $A$ and $B$. $\Re{(A)}$ and $\Im{(A)}$ are the real and imaginary part of scalar/vector/matrix $A$. $\mathbf I_{N}$ represents an identity matrix for dimension $N \times N$. $A*B$ represents the linear convolution of $A$ and $B$. $\mathbf A^{\downarrow l}$ represents $l$ sample delayed version of the vector $\mathbf A$ with zero padding at the front. We use $\{\bar{.}\}$ or $\{\tilde{.}\}$ over any variable to represent the real and imaginary part of that scalar/vector/matrix respectively.
\begin{figure*}[t]
	\centering
	\includegraphics[scale=0.32]{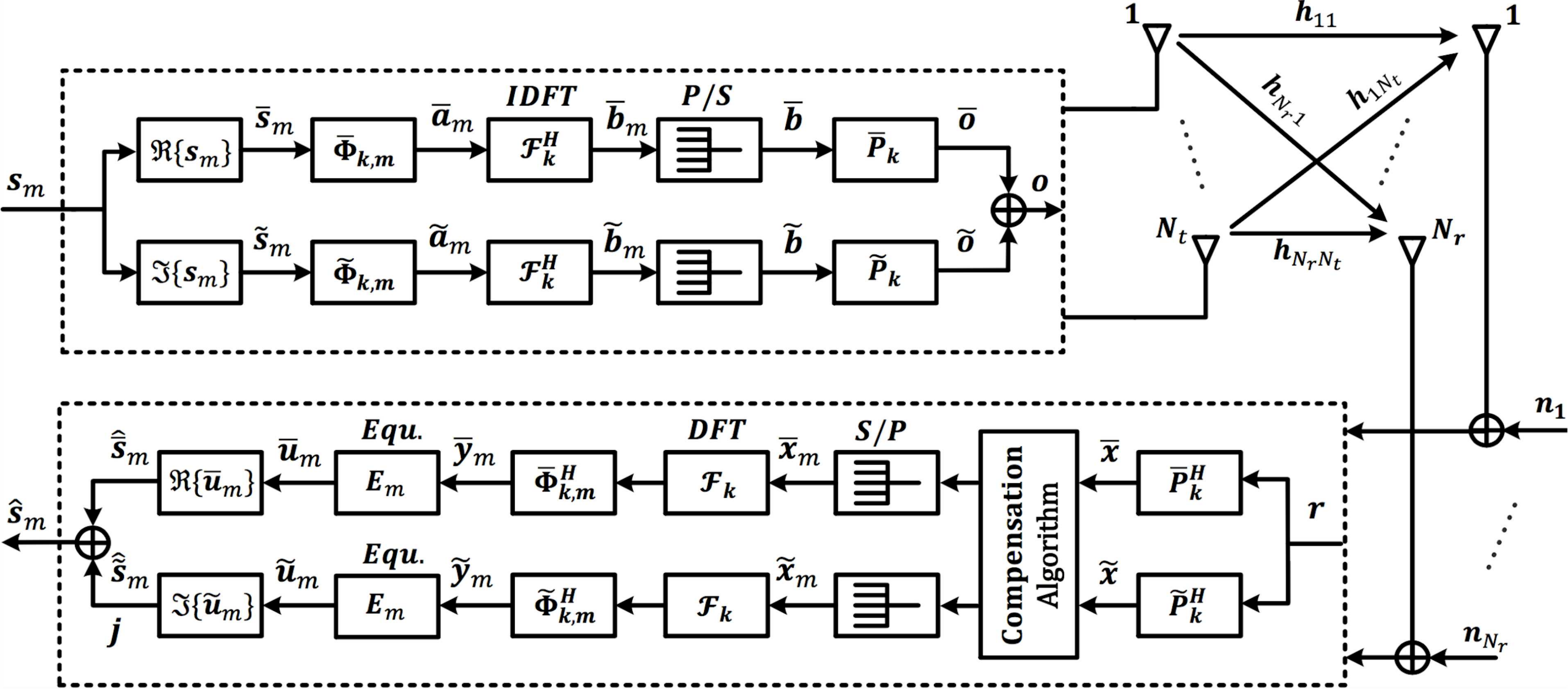}
	\caption{Blocks diagrams for MIMO-FBMC transmitter and receiver in matrix operation form}
	\label{fig:Tx_Rx}
\end{figure*}
\section{Problem Formulation}
\subsection{MIMO-FBMC System Model}
In our analysis of MIMO-FBMC system, we assumed $N_t$ transmit antennas are used to transmit multiple FBMC signals simultaneously which are received by $N_r$ received antennas, where $N_t\le N_r$. The block diagram for both transmitter and receiver of a MIMO-FBMC is shown in Fig. \ref{fig:Tx_Rx}, where real and imaginary branches i.e. $\mathcal{I}$ and $\mathcal{Q}$ branches are processed simultaneously and independently. The MIMO-FBMC system model follows a block based processing approach where each block contains $M$ FBMC symbols with each symbol containing $N$ subcarriers in frequency domain. Therefore, we can represent each MIMO-FBMC block as ${\bf{S}}=[{\mathbf{s}_0},{\mathbf{s}_1},{\mathbf{\cdots}},{\mathbf{s}_{M-1}}]\in \mathbb{C}^{NN_t\times M}$ where ${\mathbf{s}_m}=[{\mathbf{s}_{m,0}},{\mathbf{s}_{m,1}},{\mathbf{\cdots}},{\mathbf{s}_{m,N-1}}]^T\in \mathbb{C}^{NN_t\times1}$. The transmitted signal on the $n^{th}$ subcarrier in a MIMO system is an $N_t\times1$ vector, i.e., ${\mathbf{s}_{m,n}}=[s_{m,n,1},s_{m,n,2},\cdots,s_{m,n,N_t}]^T\in \mathbb{C}^{N_t\times1}$. Each $s_{m,n,j}$ represents a complex signal on $n^{th}$ subcarrier of $m^{th}$ FBMC symbol transmitted by $j^{th}$ transmitting antenna. Hence, $MNN_t$ QAM symbols are transmitted in one FBMC block. Note that a precoding scheme such as ZF can be applied at the transmitter side when the channel state information (CSI) is available. In such cases the performance of a system can be further enhanced. However, the focus of this paper is to analyze the performance of MIMO-FBMC system with finite filter length and FOT. Therefore, the analysis presented in Section \ref{sec4} is based on unitary precoding matrix but is easily extend-able to the precoding case as well. Moreover, the power of modulated symbols $s_{m,n,j}$ is represented as $\delta^2$ i.e. $\mathbb{E}\{\|s_{m,n,j}\|^2\}= \delta^2$. The real and imaginary parts of $\mathbf{s}_m$ are represented as $\mathbf{\bar{s}}_m$  and $\mathbf{\tilde{s}}_m$ respectively.
\subsection{MIMO Channel Impulse Response}
We assume the system operates over a slowly-varying fading channel i.e. quasi-static fading channel. In such a scenario, it is plausible to assume that the duration of each transmitted data block is smaller than the coherence time of the channel, therefore, the random fading coefficients stay constant over the duration of each block \cite{tse2005fundamentals}. In this case, we define the multipath channel as a $l$-tap channel impulse response (CIR) matrix with the $l^{th}$-tap power being ${\rho^2_l}$. It is also assumed that the average power remains constant during transmission of whole block. The CIR matrix $\mathbf H$ is defined as
\begin{eqnarray}\label{eq:1}
\mathbf H &=& [\mathbf H_{0},\mathbf H_{1},\cdots,\mathbf H_{L-1}]^T \nonumber\\ 
&=& [\rho_0 \mathbf Z_{0},\rho_1 \mathbf Z_{1},\cdots,\rho_{L-1} \mathbf Z_{L-1}]^T
\end{eqnarray}
where $\mathbf H_{l}$ defines the $l^{th}$ matrix valued CIR coefficient of the channel between all the antennas and is represented as
\begin{eqnarray}\label{eq:2}
\mathbf H_{l}=\rho_l \mathbf Z_{l} = \rho_l 
\begin{bmatrix}
\mathbf z_{11}(l)  &  \cdots & \mathbf z_{1N_t}(l)\\
\vdots  &  \ddots & \vdots\\
\mathbf z_{N_{r}1}(l)  &  \cdots & \mathbf z_{N_{r}N_{t}}(l)\\
 \end{bmatrix}
 \in \mathbb{C}^{N_r\times N_t}
\end{eqnarray}
\indent The random variable $\mathbf z_{ij}(l)$ with complex Gaussian distribution as $\mathcal{CN}(0,1)$ represents the multipath fading factor for $l^{th}$ tap of the quasi-static rayleigh fading channel between $j^{th}$ transmit antenna and $i^{th}$ receive antenna. Note that we consider co-located transmit and receive antennas to simplify our analysis. However, if we consider either transmit or receive antennas to be geographically separated, the analysis can easily be extended by considering the common coefficient $\rho_l$ to be different among the antennas. 
\vspace*{-0.2cm}
\subsection{Prototype Filters / Filter Matrices}
Ideally, an infinite filter length $(K=\infty)$ is required to provide the best performance. However, a finite filter length (e.g. overlapping factor $K=4\sim6$) is used in practice in a FBMC system to achieve comparable system performance. To generalize the derivation, the filter overlapping factor is taken as $K$, therefore, $KN$ is the total length of the prototype filter i.e. ${\mathbf{\bar{w}}}=[{\mathbf{\bar{w}}_0},{\mathbf{\bar{w}}_1},\cdots,{\mathbf{\bar{w}}_{K-1}}]=[\bar{w}_0,\bar{w}_0,\cdots,\bar{w}_{KN-1}]\in \mathbb{R}^{1\times KN}$. The $\mathcal{I}$ branch filter matrix $\mathbf {\bar P}_{orig} \in \mathbb{R}^{(K+M-1)NN_t \times MNN_t}$ can be expressed as
\begin{eqnarray}\label{eq:3}
\!\!\!\!\!\!\mathbf {\bar P}_{orig} \!=\!\!
\begin{bmatrix}
\begin{matrix}
\\
\mathbf {\bar P}_{i_F}\\
\\
\hdashline[0.2pt/2pt]
\\
\mathbf {\bar P}\\
\\
\hdashline[0.2pt/2pt]
\\
\mathbf {\bar P}_{i_R}
\\
\end{matrix}
\end{bmatrix}
\!\!=\!\!
\begin{bmatrix}
\begin{matrix}
\mathbf {\bar W}_0 \!\!\!&\!\!\! \mathbf{0} \!\!\!&\!\!\! \mathbf{0} \!\!\!&\! \cdots \!&\!\!\! \mathbf{0}\!\!&\!\!\!\!\! \mathbf{0}\!\!\\ 
\mathbf {\bar W}_1 \!\!\!&\!\!\! \mathbf {\bar W}_0 \!\!\!&\!\!\! \mathbf{0} \!\!\!&\! \cdots \!&\!\!\! \mathbf{0}\!\!&\!\!\! \!\!\mathbf{0}\!\!\\ 
\vdots \!\!\!&\!\!\! \vdots \!\!\!&\!\!\! \mathbf {\bar W}_0 \!\!\!&\! \cdots \!&\!\!\! \mathbf {0}\!\!&\!\!\! \!\!\mathbf {0}\!\!\\
\mathbf {\bar W}_{t-1}  \!\!\!&\!\!\! \mathbf {\bar W}_{t-2} \!\!\!&\!\!\! \vdots \!\!\!&\! \cdots \!&\!\!\! \vdots\!\!&\!\!\! \!\!\vdots\!\!\\
\hdashline[0.2pt/2pt]
\mathbf {\bar W}_{t} \!\!\!&\!\!\! \mathbf {\bar W}_{t-1} \!\!\!&\!\!\! \mathbf {\bar W}_{t-2} \!\!\!&\! \cdots \!&\!\!\! \mathbf {0}\!\!&\!\!\! \!\!\mathbf {0}\!\!\\ 
\mathbf {\bar W}_{t+1} \!\!\!&\!\!\! \mathbf {\bar W}_t \!\!\!&\!\!\! \mathbf {\bar W}_{t-1} \!\!\!&\! \cdots \!&\!\!\! \mathbf {\bar W}_{0}\!\!&\!\!\!\!\! \mathbf {0}\!\!\\ 
\vdots \!\!\!&\!\!\! \mathbf {\bar W}_{t+1} \!\!\!&\!\!\! \mathbf {\bar W}_t \!\!\!&\! \cdots \!&\!\!\! \mathbf {\bar W}_{1}\!\!&\!\!\!\!\! \mathbf {\bar W}_{0}\!\!\\ 
\mathbf {\bar W}_{K-1} \!\!\!&\!\!\! \vdots \!\!\!&\!\!\! \mathbf {\bar W}_{t+1} \!\!\!&\! \cdots \!&\!\!\! \vdots\!\!&\!\!\!\!\! \mathbf {\bar W}_{1}\!\!\\ 
\mathbf{0} \!\!\!&\!\!\! \mathbf {\bar W}_{K-1} \!\!\!&\!\!\! \vdots \!\!\!&\! \cdots \!&\!\!\! \mathbf {\bar W}_{K-t}\!\!&\!\!\!\!\! \vdots\!\!\\ 
\mathbf{0} \!\!\!&\!\!\! \mathbf{0} \!\!\!&\!\!\! \mathbf {\bar W}_{K-1} \!\!\!&\! \cdots \!&\!\!\! \vdots\!\!&\!\!\!\!\! \mathbf {\bar W}_{K-t}\!\!\\
\hdashline[0.2pt/2pt]
\vdots \!\!\!&\!\!\! \vdots \!\!\!&\!\!\! \vdots \!\!\!&\! \cdots \!&\!\!\! \mathbf {\bar W}_{K-1} \!\!&\!\!\!\!\! \vdots \!\!\\
\mathbf{0} \!\!\!&\!\!\! \mathbf{0} \!\!\!&\!\!\! \mathbf{0} \!\!\!&\! \cdots \!&\!\!\! \mathbf {0}\!\!&\!\!\!\!\! \mathbf {\bar W}_{K-1}\!
\end{matrix}
\end{bmatrix}
\!\!\!\!\!\!\!\!
\end{eqnarray}
where, $\mathbf{\bar{W}}_k=diag(\mathbf{\bar{w}}_k)\in \mathbb{R}^{N\times N}$ for $k=0,1,2,\cdots,K-1$ and $\mathbf{\bar{w}}_k=[\bar{w}_{kN},\bar{w}_{kN+1}, \cdots,\bar{w}_{kN+N-1}]\in \mathbb{R}^{1\times N}$ while $t = \lfloor \frac{K}{2} \rfloor$. The value of $t$ defines the truncated matrix $\bar{\mathbf P}$ as shown in (\ref{eq:3}). The prototype filter matrix for the $\mathcal{Q}$ branch is defined in the same manner. The only difference is that the $\mathcal{Q}$ branch filter is a shifted version of the $\mathcal{I}$ branch filter i.e. ${\mathbf{\tilde{w}}}=[{\mathbf{\tilde{w}}_0},{\mathbf{\tilde{w}}_1},\cdots,{\mathbf{\tilde{w}}_{K-1}}]=[\tilde{w}_0,\tilde{w}_0,\cdots,\tilde{w}_{KN-1}]=[\bar{w}_{\frac{N}{2}},\bar{w}_{\frac{N}{2}+1},\cdots,\bar{w}_{KN-1},\bar{w}_0, \bar{w}_1,\cdots,\tilde{w}_{\frac{N}{2}-1}]\in \mathbb{R}^{1\times KN}$. Shifting prototype filter in the $\mathcal{Q}$ branch instead of offsetting the QAM symbols makes the overall design simpler. Similarly, the $\mathcal{Q}$ branch truncated filter matrix $\mathbf{\tilde{P}}$ is defined in the same manner as described for the $\mathcal{I}$ branch with $\mathbf{\tilde{W}}_k=\textrm{diag}(\mathbf{\tilde{w}}_k)\in \mathbb{R}^{N\times N}$ for $k=0,1,2,\cdots,K-1$ and $\mathbf{\tilde{w}}_k=[\tilde{w}_{kN},\bar{w}_{kN+1},\cdots,\tilde{w}_{kN+N-1}]\in \mathbb{R}^{1\times N}$.
\vspace*{-0.1cm}
\section{MIMO-FBMC Matrix Model}
The MIMO-FBMC matrix model is derived by extending our previous work on a SISO-FBMC system \cite{7549072}. It is worth mentioning that the derivation of MIMO-FBMC matrix model is not a simple SISO to MIMO mapping. Signal definition, transmit processing, channel modeling, as well as receive processing including channel equalization has to be redefined. The derived MIMO-FBMC model also incorporates FOT as well as the proposed compensation algorithm at the receiver.
\vspace*{-0.3cm}
\subsection{Transmit Processing}
We will only focus on the real branch in detail since the imaginary branch will follow the same procedure.\vspace*{.1cm}
\subsubsection{\textit{Real Branch Processing}}
According to  Fig. \ref{fig:Tx_Rx}, the signal $\mathbf{\bar{s}}_m$ is first multiplied by a phase shifter matrix $\mathbf{\bar \Phi}_m$ symbol by symbol i.e., \vspace*{-.3cm}
\begin{eqnarray}\label{eq:4}
{\mathbf{\bar{a}}_m}&=&({\mathbf{\bar \Phi}_m}\otimes {\mathbf{I}_{N_t}}){\mathbf{\bar{s}}_m}\nonumber \\
&=& \mathbf{\bar \Phi}_{k,m}{\mathbf{\bar{s}}_m}\in \mathbb{C}^{NN_t\times1}
\end{eqnarray}
where $\mathbf{\bar \Phi}_m$ is a diagonal matrix i.e. ${\mathbf{\bar \Phi}_m}=\textrm{diag}[e^{\frac{-j\pi(0+2m)}{2}},e^{\frac{-j\pi(1+2m)}{2}},\cdots ,e^{\frac{-j\pi(N-1+2m)}{2}}]\in \mathbb{C}^{N\times N}$. Note that $\mathbf{\bar \Phi}_{k,m}$ represents the kronecker product ${\mathbf{\bar \Phi}_m}\otimes {\mathbf{I}_{N_t}}$ that yields a matrix of size $NN_t\times NN_t$.
\vspace*{.1cm}
\subsubsection{\textit{Real Branch IDFT Processing}}
Signal after the phase shifter matrix will pass through an $N$ point IDFT (inverse discrete Fourier transform) block ${\mathbf{\mathcal{F}}^H}$ i.e.
\begin{eqnarray}\label{eq:5}
{\mathbf{\bar{b}}_m} &=& ({\mathbf{\mathcal{F}}^H}\otimes {\mathbf{I}_{N_t}}){\mathbf{\bar{a}}_m} \nonumber \\
&=&\mathbf{\mathcal{F}}_k^H{\mathbf{\bar{a}}_m}\in \mathbb{C}^{NN_t\times 1}
\end{eqnarray}
where $\mathbf{\mathcal{F}}_k^H = {\mathbf{\mathcal{F}}^H}\otimes {\mathbf{I}_{N_t}}$. Signal after the IDFT block can be represented as ${\mathbf{\bar{b}}}=[{\mathbf{\bar{b}}_0};{\mathbf{\bar{b}}_1};{\mathbf{\cdots}};{\mathbf{\bar{b}}_{M-1}}]=[{\mathbf{\mathcal{F}}_k^H}{\mathbf{\bar{a}}_0};{\mathbf{\mathcal{F}}_k^H}{\mathbf{\bar{a}}_1};{\mathbf{\cdots}};{\mathbf{\mathcal{F}}_k^H}{\mathbf{\bar{a}}_{M-1}}]\in \mathbb{C}^{MNN_t\times1}$. Here IDFT is a block wise operation since each modulated subcarrier is a column vector of size $N_t\times1$ and ${\mathbf{\mathcal{F}}_k^H}$ is a generalized $NN_t$ point IDFT matrix. \vspace*{.1cm}
\subsubsection{\textit{Real Branch Prototype Filter}}
The signal is then passed through a prototype filter in $\mathcal{I}$ and $\mathcal{Q}$ branches independently. In general, prototype filters are linearly convolved with the input signal. In order to represent a complete system in matrix form we have defined a prototype filter matrix  $\mathbf{\bar{P}}$ in a manner that when this filter matrix is multiplied by vector $\mathbf{\bar{b}}$; the multiplication of matrices is equivalent to the required linear convolution process. The output of the $\mathcal{I}$ branch filter can be written as\vspace*{-.3cm}
\begin{eqnarray}\label{eq:6}
\mathbf{\bar{o}} &=& \mathbf{\bar{P}}_{k,orig}\mathbf{\bar{b}}
\end{eqnarray}
where $\mathbf{\bar{P}}_{k,orig} = \mathbf{\bar{P}}_{orig}\otimes \mathbf I_{N_t}$. Note that the output of the real branch filter $\mathbf{\bar{o}}$ has $(K-1)NN_t$ more samples than the input due to the linear convolution process. Hence, to keep the orthogonality (minimum interference from other subcarriers and symbols), all of these samples have to be transmitted to the receiver side. However, the transmission efficiency $\eta$ will drop by the proportion of 
\vspace*{-0.1cm}
\begin{eqnarray}\label{eq:7}
\eta &=& \frac{M}{K+M-1}
\end{eqnarray}
\indent It can be seen from (\ref{eq:7}), that transmission efficiency $\eta$ is high only if $M$ is large. Another way to achieve higher $\eta$ is to truncate $\mathbf {\bar P}_{orig}$ to improve the spectral efficiency. However, truncation may lead to interferences in the system that can significantly degrade the system performance. Without any compensation, the maximum allowable cut off symbols would be $K-2$ so as to keep the signals detectable \cite{7390479}. However, with compensation we can completely discard all the $K-1$ symbols while still keeping the signals detectable. The truncation should take place at the first $i_F$ and the last $i_R$ rows of $\mathbf {\bar P}_{orig}$ such that $i_F+i_R \le K-1$ as shown in (\ref{eq:3}), where $\mathbf{\bar P}_{i_F}$ is first $i_FN$ rows and $\mathbf{\bar P}_{i_R}$ is the last $i_RN$ rows of $\mathbf{\bar P}_{orig}$. The middle part of $\mathbf{\bar P}_{orig}$ i.e. $\mathbf{\bar P}$ which is the truncated filter matrix will be used at transmitter side to improve the spectral efficiency of the system. The performance loss due to the truncation of $\mathbf{\bar P}_{orig}$ will be compensated at the receiver side and is discussed later in Section \ref{sec4}. The output of real branch truncated filter can be written as
\begin{eqnarray}\label{eq:8}
\mathbf{\bar{o}}&=&(\mathbf{\bar{P}}\otimes \mathbf I_{N_t})\mathbf{\bar{b}}\nonumber \\ 
&=&\mathbf{\bar{P}}_k\mathbf{\bar{b}} \in \mathbb{C}^{MNN_t\times1}
\end{eqnarray}
\subsubsection{\textit{Imaginary Branch Processing Including Prototype Filtering}}
Similar process is followed for the $\mathcal{Q}$ branch i.e. the signal $\mathbf{\tilde{s}}_m$ is first multiplied by a phase shifter matrix $\mathbf{\tilde \Phi}_m=j\mathbf{\bar \Phi}_m$ symbol by symbol i.e., 
\begin{eqnarray}\label{eq:9}
\mathbf{\tilde{a}}_m&=&({\mathbf{\tilde \Phi}_m}\otimes {\mathbf{I}_{N_t}}){\mathbf{\tilde{s}}_m}\nonumber \\
&=&{\mathbf{\tilde \Phi}}_{k,m}{\mathbf{\tilde{s}}_m}\in \mathbb{C}^{NN_t\times1}
\end{eqnarray}
After the phase shifter matrix, the signal will pass through an $N$ point IDFT block $\mathcal{F}^H$ as 
\begin{eqnarray}\label{eq:10}
{\mathbf{\tilde{b}}_m}&=& ({\mathbf{\mathcal{F}}^H}\otimes {\mathbf{I}_{N_t}}){\mathbf{\tilde{a}}_m}\nonumber \\
&=& \mathbf{\mathcal{F}}_k^H{\mathbf{\tilde{a}}_m}\in \mathbb{C}^{NN_t\times 1}
\end{eqnarray}
The signal after IDFT block can be represented as ${\mathbf{\tilde{b}}}=[{\mathbf{\tilde{b}}_0};{\mathbf{\tilde{b}}_1};{\mathbf{\cdots}};{\mathbf{\tilde{b}}_{M-1}}]=[{\mathbf{\mathcal{F}}_k^H}{\mathbf{\tilde{a}}_0};{\mathbf{\mathcal{F}}_k^H}{\mathbf{\tilde{a}}_1};{\mathbf{\cdots}};{\mathbf{\mathcal{F}}_k^H}{\mathbf{\tilde{a}}_{M-1}}]\in \mathbb{C}^{MNN_t\times1}$. Likewise, the following matrix multiplication of truncated filter matrix $\mathbf{\tilde{P}}$ and the signal vector $\mathbf{\tilde{b}}$ represents the linear convolution of the imaginary branch prototype filter and the imaginary branch input signal.
\begin{eqnarray}\label{eq:11}
\mathbf{\tilde{o}}&=&(\mathbf{\tilde{P}}\otimes \mathbf I_{N_t})\mathbf{\tilde{b}}\nonumber \\ 
&=&\mathbf{\tilde{P}}_k\mathbf{\tilde{b}} \in \mathbb{C}^{MNN_t\times1}
\end{eqnarray}
\subsection{Passing through the Channel}
The real and imaginary branch signals $\mathbf{\bar{o}}$ and $\mathbf{\tilde{o}}$ after the prototype filtering are added together and is then passed through the channel $\mathbf H$. The received signal is now represented as
\begin{eqnarray}\label{eq:12}
\mathbf  {r} &=& \mathbf H * (\mathbf {\bar o} + \mathbf {\tilde o}) + \mathbf n
\end{eqnarray}
where $\mathbf{n}=[\mathbf{n}_1,\mathbf{n}_2,\cdots,\mathbf{n}_{N_r}]^T\in \mathbb{C}^{MNN_r\times1}$ is a Gaussian noise vector with each element having zero mean and variance $\sigma^2$. To represent the convolution process given in (\ref{eq:12}) as matrix multiplication, we define the $l^{th}$ tap multipath fading factor $\mathbf Z_l$ of the MIMO channel as a block diagonal matrix by taking the kronecker product of an identity matrix $\mathbf I_{(K+M-1)N}$ with $\mathbf Z_{l}$ as
\begin{eqnarray}\label{eq:13}
\mathbf{Z}_{l,blk}=\mathbf I_{(K+M-1)} \otimes \mathbf Z_{k,l}
\end{eqnarray}
where $\mathbf Z_{k,l}=\mathbf I_{N} \otimes \mathbf Z_{l}\in\mathbb{C}^{NN_r\times NN_t}$. The block diagonal matrix $\mathbf{Z}_{l,blk}\in\mathbb{C}^{(K+M-1)NN_r\times (K+M-1)NN_t}$ has $\mathbf Z_l$ as its diagonal sub matrices. The definition of $\mathbf{Z}_{l,blk}$ implies that each FBMC symbol in a block experiences the same channel i.e. $\mathbf{Z}_{l}$ . Hence, we can write (\ref{eq:12}) as
\begin{eqnarray}\label{eq:14}
\mathbf  {r} =\sum_{l=0}^{L-1} \rho_l \mathbf Z_{l,blk}(\mathbf {\bar o}^{\downarrow N_tl} + \mathbf {\tilde o}^{\downarrow N_tl}) + \mathbf n
\end{eqnarray}
where $\mathbf {\bar o}^{\downarrow N_tl}$, $\mathbf {\tilde o}^{\downarrow N_tl}$ represents $N_t l$ samples delayed version of $\mathbf{\bar{o}}$ and $\mathbf{\tilde{o}}$ with zero padding in the front i.e. $\mathbf {\bar o}^{\downarrow N_tl}=[\mathbf 0_{N_t l\times 1};\mathbf{\bar{o}}_{q,N_t l}]$ and $\mathbf {\tilde o}^{\downarrow N_tl}=[\mathbf 0_{N_t l\times 1};\mathbf{\tilde{o}}_{q,N_t l}]$ respectively. Note that $\mathbf{\bar{o}}_{q,N_t l}$ and $\mathbf{\tilde{o}}_{q,N_t l}$ represents the first $(K\!+\!M\!-\!1)NN_t\!-\!N_t l$ elements of $\mathbf{\bar{o}}$ and $\mathbf{\tilde{o}}$ respectively. From (\ref{eq:8}) and (\ref{eq:11}) we can write $\mathbf {\bar o}^{\downarrow N_tl}=\mathbf {\bar{P}}_k^{\downarrow N_tl}\mathbf{\bar{b}}$ and $\mathbf {\tilde o}^{\downarrow N_tl}=\mathbf {\tilde{P}}_{k}^{\downarrow N_tl}\mathbf{\tilde{b}}$, where $\mathbf {\bar{P}}_{k}^{\downarrow N_tl}=[\mathbf 0_{N_t l\times MNN_t}; \mathbf{\bar{P}}_{k,q}]$ and $\mathbf {\tilde{P}}_{k}^{\downarrow N_tl}=[\mathbf 0_{N_t l\times MNN_t}; \mathbf {\tilde{P}}_{k,q}]$. Here $\mathbf{\bar{P}}_{k,q}$ and $\mathbf {\tilde{P}}_{k,q}$ are the first $(K+M-1)NN_t-N_t l$ rows of $\mathbf {\bar{P}}_{k}$ and $\mathbf {\tilde{P}}_{k}$ respectively. Eq (\ref{eq:14}) can thus be reformed as
\begin{eqnarray}\label{eq:15}
\mathbf  {r} =\sum_{l=0}^{L-1} \rho_l \mathbf Z_{l,blk}(\mathbf {\bar{P}}_{k}^{\downarrow N_tl}\mathbf{\bar{b}} + \mathbf {\tilde{P}}_{k}^{\downarrow N_tl}\mathbf{\tilde{b}})+ \mathbf n
\end{eqnarray}
\indent The above equation indicates that the truncated filter matrix $\mathbf {\bar{P}}_{k}$ and $\mathbf {\tilde{P}}_{k}$ are distorted because of the channel multipath effect and are represented as $\mathbf {\bar{P}}_{k}^{\downarrow N_tl}$ and $\mathbf {\tilde{P}}_{k}^{\downarrow N_tl}$ respectively.\\
To represent (\ref{eq:15}) in a point-wise multiplication form in frequency domain, we apply the circular convolution property by first introducing a block diagonal exchange matrix $\mathbf X_{N_t l} \in \mathbb{R}^{MNN_t\times MNN_t}$ as
\begin{eqnarray}\label{eq:16}
\mathbf {X}_{N_tl} =
 \begin{bmatrix}
  \mathbf {X}_{sub,N_tl} & \mathbf {0} & \cdots & \mathbf {0} \\
  \mathbf {0} & \mathbf {X}_{sub,N_tl} & \cdots & \mathbf {0} \\
  \vdots  & \vdots  & \ddots & \vdots  \\
  \mathbf {0} & \mathbf {0} & \cdots & \mathbf {X}_{sub,N_tl}
 \end{bmatrix}
\end{eqnarray}
with
\begin{eqnarray}\label{eq:17}
\mathbf {X}_{sub,N_tl} =
 \begin{bmatrix}
  \mathbf {0}_{N_tl\times (NN_t-N_tl)} & \mathbf {I}_{N_tl\times N_tl}  \\
  \mathbf {I}_{(NN_t-N_tl)\times (NN_t-N_tl)} & \mathbf {0}_{(NN_t-N_tl)\times N_tl}
 \end{bmatrix}
 \end{eqnarray}
As $\mathbf {X}^T_{N_tl}\mathbf {X}_{N_tl} = \mathbf I$, we have
\begin{eqnarray}\label{eq:18}
\mathbf{\bar o}^{\downarrow N_tl} = \mathbf {\bar{P}}_{k}^{\downarrow N_tl}\mathbf{\bar b}=\mathbf {\bar{P}}_{k}^{\downarrow N_tl}\mathbf {X}^T_{N_tl}\mathbf {X}_{N_tl} \mathbf{\bar b} = \mathbf{\bar{P}}_{k,e}^{\downarrow N_tl}\mathbf{\bar b}^{\downarrow N_tl}_e
\end{eqnarray}
\indent The matrix $\mathbf {X}^T_{N_tl}$ and $\mathbf {X}_{N_tl}$ are used to exchange the locations of elements of $\mathbf{\bar{P}}_{k}^{\downarrow N_tl}$ and $\mathbf{\bar b}$ respectively, such that $\mathbf{\bar{P}}_{k,e}^{\downarrow N_tl} = \mathbf{\bar{P}}_{k}^{\downarrow N_tl}\mathbf {X}^T_{N_tl}$ and $\mathbf{\bar b}^{\downarrow N_tl}_e = \mathbf {X}_{N_tl} \mathbf{\bar b}$. By multiplying the matrix $\mathbf {X}_{N_tl}$ with $\mathbf{\bar{b}}$, the last $N_tl$ symbols of its each sub-vector $\mathbf{\bar b}_m$ will be moved to the front, i.e.,
\begin{eqnarray}\label{eq:19}
\mathbf{\bar b}_{e,m}^{\downarrow N_tl} = [\mathbf b_{m,NN_t-N_tl}\cdots,\mathbf b_{m,NN_t-1},\mathbf b_{m,0},\cdots,\nonumber \\
\mathbf b_{m,NN_t-N_tl-1}]^T \in \mathbb{C}^{NN_t\times 1}
\end{eqnarray}
Likewise,
\begin{eqnarray}\label{eq:20}
\mathbf{\bar b}_e^{\downarrow N_tl} = [\mathbf{\bar b}_{e,0}^{\downarrow N_tl};\mathbf{\bar b}_{e,1}^{\downarrow N_tl};\cdots;\mathbf{\bar b}_{e,M-1}^{\downarrow N_tl}] \in \mathbb{C}^{MNN_t\times 1}
\end{eqnarray}
The effect is similar when multiplying $\mathbf {X}^T_{N_tl}$ with $\mathbf{\bar{P}}_{k}^{\downarrow N_tl}$. $\mathbf {X}^T_{N_tl}$ only changes the elements location in $\mathbf{\bar{P}}_{k}^{\downarrow N_tl}$.
Similarly, we can write $\mathbf {\tilde o}^{\downarrow N_tl}$ as
\begin{eqnarray}\label{eq:21}
\mathbf{\tilde o}^{\downarrow N_tl} = \mathbf {\tilde{P}}_{k}^{\downarrow N_tl}\mathbf{\tilde b}=\mathbf {\tilde{P}}_{k}^{\downarrow N_tl}\mathbf {X}^T_{N_tl}\mathbf {X}_{N_tl} \mathbf{\tilde b} = \mathbf{\tilde{P}}_{k,e}^{\downarrow N_tl}\mathbf{\tilde b}^{\downarrow N_tl}_e
\end{eqnarray}
Substituting (\ref{eq:18}) and (\ref{eq:21}) into (\ref{eq:15}) yields
\begin{eqnarray}\label{eq:22}
\mathbf  {r} =\sum_{l=0}^{L-1} \rho_l \mathbf Z_{l,blk}(\mathbf{\bar{P}}_{k,e}^{\downarrow N_tl}\mathbf{\bar b}^{\downarrow N_tl}_e + \mathbf{\tilde{P}}_{k,e}^{\downarrow N_tl}\mathbf{\tilde b}^{\downarrow N_tl}_e) + \mathbf n
\end{eqnarray}
\indent It can be observed that the non zero elements of $\mathbf{\bar{P}}_{k,e}^{\downarrow N_tl}$ and  $\mathbf{\bar{P}}_{k}$ are very close i.e. the nonzero elements of $\mathbf{\bar{P}}_{k,e}^{\downarrow N_tl}$ are only delayed by $N_tl$ elements as compared to the elements in $\mathbf{\bar{P}}_{k}$. If the non-zero $i^{th}$ row and $k^{th}$ column element of $\mathbf{\bar{P}}_{k}$ is $w_{n}$, then the element of $\mathbf{\bar{P}}_{k,e}^{\downarrow N_tl}$ at the same location will be $w_{n+N_tl}$. Since $N\gg L$, the difference between $w_{n}$ and $w_{n+N_t l}$ is very small as the adjacent elements of the prototype filter are close to each other. Eq (\ref{eq:22}) can thus be written as
\begin{eqnarray}\label{eq:23}
\mathbf  {r} &\approx& \sum_{l=0}^{L-1} \rho_l \mathbf Z_{l,blk}(\mathbf{\bar{P}}_{k}\mathbf{\bar b}^{\downarrow N_tl}_e + \mathbf{\tilde{P}}_{k}\mathbf{\tilde b}^{\downarrow N_tl}_e)+ \mathbf n
\end{eqnarray}
\subsection{Receive Processing}
On the receiver side, the signal $\mathbf r$ is received by $N_r$ received antennas and is fed to the real and imaginary branches of the receiver as shown in Fig. \ref{fig:Tx_Rx} for independent processing.\vspace*{.1cm}
\subsubsection{\textit{Real Branch Processing}}
Following the similar approach, we will focus on the real branch processing and the imaginary branch processing follows the same procedure. In the real branch, signals from $N_r$ received antennas are passed through the real branch received filters leading to the following output
\begin{eqnarray}\label{eq:24}
\mathbf  {\bar x} &=&\mathbf{\bar{P}}_{k}^{H}\mathbf r \nonumber\\
&=&\mathbf{\bar{P}}_{k}^{H}\mathbf{\bar{P}}_{k}\sum_{l=0}^{L-1} \rho_l \mathbf Z_{l,blk}\mathbf{\bar b}^{\downarrow N_tl}_e + \mathbf{\bar{P}}_{k}^{H}\mathbf{\tilde{P}}_{k}\sum_{l=0}^{L-1} \rho_l\mathbf Z_{l,blk} \nonumber\\
&&\mathbf{\tilde b}^{\downarrow N_tl}_e + \mathbf{\bar{P}}_{k}^{H}\mathbf n 
\end{eqnarray}
Autocorrelation and cross-correlation matrices of $\mathbf{\bar{P}}_{k}$ and $\mathbf{\tilde{P}}_{k}$ are defined as $\mathbf {\bar{\bar G}}_{k} = \mathbf{\bar{P}}_{k}^{H}\mathbf{\bar{P}}_{k}$, $\mathbf {\bar{\tilde G}}_{k} = \mathbf{\bar{P}}_{k}^{H}\mathbf{\tilde{P}}_{k}$, $\mathbf {\tilde{\tilde G}}_{k} = \mathbf{\tilde{P}}_{k}^{H}\mathbf{\tilde{P}}_{k}$ and $\mathbf {\tilde{\bar G}}_{k} = \mathbf{\tilde{P}}_{k}^{H}\mathbf{\bar{P}}_{k}$. Here $\mathbf {\bar{\bar G}}_{k}, \mathbf {\bar{\tilde G}}_{k}, \mathbf {\tilde{\tilde G}}_{k}$ and $\mathbf {\tilde{\bar G}}_{k} \in \mathbb{R}^{MNN_r \times MNN_t}$\\ 
The above equation (\ref{eq:24}) can now be written as
\begin{eqnarray}\label{eq:25}
\mathbf  {\bar x} &=&\mathbf {\bar{\bar G}}_{k}\sum_{l=0}^{L-1} \rho_l \mathbf Z_{l,blk}\mathbf{\bar b}^{\downarrow N_tl}_e + \mathbf {\bar{\tilde G}}_{k} \sum_{l=0}^{L-1} \rho_l \mathbf Z_{l,blk}\mathbf{\tilde b}^{\downarrow N_tl}_e \nonumber \\
&&+ \mathbf{\bar{P}}_{k}^{H}\mathbf n
\end{eqnarray}
\subsubsection{\textit{Real Branch DFT Processing and Phase Shifting}}
The signal vector at the output of the real branch filter matrix i.e. $\mathbf {\bar{P}}_{k}^H$ is represented as $\mathbf{\bar{x}}=[\bar{x_0},\bar{x_1},\cdots,\bar{x_{MNN_r-1}}]^T\in \mathbb{C}^{MNN_r\times 1}$ and is then passed through a serial to parallel conversion to split the vector into $M$ segments, each of which has $NN_r$ elements to perform $N$-point DFT and phase shifting process. The $m^{th}$ segment of the vector $\mathbf {\bar{x}}$ is represents as $\mathbf{\bar{x}}_m=[\bar{x}_{mNN_r},\bar{x}_{mNN_r+1},\cdots,\bar{x}_{mNN_r+NN_r-1}]^T\in \mathbb{C}^{NN_r\times1}$ for $m\in 0,1,\cdots,M-1$. The signal is now represented as $\mathbf {\bar{x}}=[\mathbf{\bar{x}}_{0},\mathbf{\bar{x}}_{1},\cdots,\mathbf{\bar{x}}_{M-1}]\in\mathbb{C}^{NN_r\times M}$ where $\mathbf {\bar{x}}_{m}=[\mathbf{\bar{x}}_{m,0},\mathbf{\bar{x}}_{m,1},\cdots,\mathbf{\bar{x}}_{m,N-1}]^T\in \mathbb{C}^{NN_r\times 1}$ in which  $\mathbf {\bar{x}}_{m,n}=[\bar{x}_{m,n,1},\cdots,\bar{x}_{m,n,N_r}]^T\in \mathbb{C}^{N_r\times 1}$. Each $\bar{x}_{m,n,i}$ represents the real signal on $n^{th}$ modulated subcarrier for $m^{th}$ FBMC symbol received by $i^{th}$ receiving antenna. Using equation (\ref{eq:25}), we can write signal vector $\mathbf{\bar{x}}_{m}$ as
\begin{eqnarray}\label{eq:26}
\mathbf  {\bar x}_m \!\!\!\!&=&\!\!\!\!\sum_{i=0}^{M-1}\bar {\bar{\mathbf G}}_{k,m,i}\sum_{l=0}^{L-1} \rho_l \mathbf Z_{k,l}\mathbf{\bar b}^{\downarrow N_tl}_{e,i} + \sum_{i=0}^{M-1}\bar {\tilde{\mathbf G}}_{k,m,i}\nonumber \\
&&\sum_{l=0}^{L-1} \rho_l \mathbf Z_{k,l}\mathbf{\tilde b}^{\downarrow N_tl}_{e,i}+ \mathbf{\bar{P}}^{H}_{k,m}\mathbf n
\end{eqnarray}
where $\bar {\bar{\mathbf G}}_{k,m,i}$ and $\bar {\tilde{\mathbf G}}_{k,m,i}$ are the $m^{th}$ row and $i^{th}$ column sub-matrices of $\mathbf {\bar{\bar G}}_k$ and $\mathbf {\bar{\tilde G}}_k$ respectively.
The signal after DFT and phase shifting is represented as
\begin{eqnarray}\label{eq:27}
\mathbf {\bar{y}}_m = {\mathbf{\bar \Phi}}_{k,m}^H\mathbf {\mathcal{F}}_{k}\mathbf {\bar{x}}_m
\end{eqnarray}
where ${\mathbf{\bar \Phi}}_{k,m}^H={\mathbf{\bar \Phi}}_{m}^H\otimes \mathbf I_{N_r}$ and $\mathbf {\mathcal{F}}_{k}=\mathbf {\mathcal{F}}\otimes \mathbf I_{N_r}\in \mathbb{C}^{NN_r\times NN_r}$. Hence, $\mathbf {\bar{y}}_m\in \mathbb{C}^{NN_r\times 1}$ can now be simplified by substituting (\ref{eq:26}) into (\ref{eq:27}) as follows before the channel equalization.
\begin{eqnarray}\label{eq:28}
\mathbf  {\bar y}_m &=& \underbrace{{\mathbf{\bar \Phi}}_{k,m}^H\mathbf {\mathcal{F}}_{k}\sum_{i=0}^{M-1}
\bar{\bar{\mathbf G}}_{k,m,i}\sum_{l=0}^{L-1}\rho_l\mathbf Z_{k,l}\mathbf{\bar b}^{\downarrow N_tl}_{e,i}}_{\mathbf  {{\bar u}}_{R,m}}\nonumber \\
&&+\underbrace{{\mathbf{\bar \Phi}}_{k,m}^H\mathbf {\mathcal{F}}_{k}\sum_{i=0}^{M-1}\bar{\tilde {\mathbf G}}_{k,m,i}\sum_{l=0}^{L-1}\rho_l\mathbf Z_{k,l} \mathbf{\tilde b}^{\downarrow N_tl}_{e,i}}_{{\mathbf  {\bar u}_{I,m}}}\nonumber\\
&&+ {\mathbf{\bar \Phi}}_{k,m}^H\mathbf {\mathcal{F}}_{k}\mathbf{\bar{P}}^{H}_{k,m}\mathbf{n}
\end{eqnarray}
\indent In (\ref{eq:28}), the third term is the noise processed by the prototype filter, DFT and the phase shifter. The term $\mathbf{\bar u}_{I,m}$ is the interference caused by the imaginary part of the signal ($\mathbf {\tilde s}_m$). Whereas, the first term $\mathbf{\bar u}_{R,m}$ contains the actual desired symbol ($\mathbf {\bar s}_m$). In $\mathbf{\bar u}_{R,m}$, we can write $\sum_{l=0}^{L-1}\rho_l\mathbf Z_{k,l} \mathbf{\bar b}^{\downarrow N_tl}_{e,i} = \mathbf H_{cir}\mathbf {\bar b}_i$. The matrix $\mathbf {H}_{cir}$ is an $NN_r \times NN_t$ block circulant matrix. In general, an $NN_r \times NN_t$ block circulant matrix is fully defined by its first $NN_r \times N_t$ block matrices. In our case, $\mathbf H_{cir}$ is determined by $[\mathbf H_{0},\mathbf H_{1},\cdots,\mathbf H_{L-1},\mathbf 0_{(N-L)N_r \times N_t}]^T\in \mathbb{C}^{NN_r \times N_t}$ 
\begin{eqnarray}\label{eq:29}
\mathbf  {\bar u}_{R,m} \!\!\!\!\! &=&\!\!\!\!\!{\mathbf{\bar \Phi}}_{k,m}^H\mathcal{F}_k \big[\!\!\sum_{i=0}^{M-1}\!\!\!\bar{\bar{\mathbf G}}_{k,m,i}\mathbf{H}_{cir}\mathbf {\bar b}_i\big]\nonumber\\
\!\!\!\!\!&=&\!\!\!\!\!{\mathbf{\bar \Phi}}_{k,m}^H\mathcal{F}_k\!\!\!\sum_{i=0}^{M-1}\!\!\!\bar{\bar{\mathbf G}}_{k,m,i} \mathcal{F}_k^H \!\!\mathcal{F}_k\mathbf{ H}_{cir}\mathcal{F}_k^H \mathcal{F}_k\mathbf {\bar b}_i
\end{eqnarray}
where $\mathcal{F}_k^H \mathcal{F}_k = \mathbf I$. Then we can use the circular convolution property as follows (pp.129-130) \cite{sesia2011lte}.
\begin{eqnarray}\label{eq:30}
\mathbf  {\bar u}_{R,m} \!\!\! \!\!\!&=&\!\!\!\!\!{\mathbf{\bar \Phi}}_{k,m}^H\mathcal{F}_k\!\!\!\sum_{i=0}^{M-1}\!\!\!\bar{\bar{\mathbf G}}_{k,m,i} \mathcal{F}_k^H \!\!\underbrace{\mathcal{F}_k\mathbf{H}_{cir}\mathcal{F}_k^H}_{\mathbf C}\mathcal{F}_k\mathbf {\bar b}_i
\end{eqnarray}
where $\mathbf C$ is the frequency domain channel coefficients in block diagonal matrix form  and is given as $\mathbf C=\textrm{blkdiag}[\mathbf C_{0},\mathbf C_{1},\cdots,\mathbf C_{N-1}] \in \mathbb{C}^{NN_r \times NN_t}$. The $n^{th}$ block diagonal element in the frequency response of the MIMO channel can be represented as $\mathbf C_{n}=\sum_{l=0}^{L-1}\mathbf H_{l}e^{-j\frac{2\pi}{N}nl}\in \mathbb{C}^{N_r \times N_t}$. $\mathcal{F}_k(\mathbf {\bar b}_{i})$ denotes the DFT processing of $\mathbf {\bar b}_{i}$ and according to (\ref{eq:5}) and (\ref{eq:4}), we have $\mathcal{F}_k(\mathbf {\bar b}_{i}) = \mathbf {\bar a}_i = \mathbf{\bar \Phi}_{k,i}\mathbf {\bar s}_i$, substituting it into (\ref{eq:30}) leads to
\begin{eqnarray}\label{eq:31}
\mathbf  {\bar u}_{R,m}\!\!\!\!\! &=&\!\!\!\!\!{\mathbf{\bar \Phi}}_{k,m}^H\mathcal{F}_k\sum_{i=0}^{M-1}\bar{\bar{\mathbf G}}_{k,m,i} \mathcal{F}_k^H \mathbf C \mathbf{\bar\Phi}_{k,i}\mathbf {\bar s}_i \nonumber \\
\!\!\!\!\!&=&\!\!\!\!\!\sum_{i=0}^{M-1}{\mathbf{\bar \Phi}}_{k,m}^H\mathcal{F}_k\bar{\bar{\mathbf G}}_{k,m,i} \mathcal{F}_k^H \mathbf{ \bar\Phi}_{k,i}\mathbf C \mathbf {\bar s}_i
\end{eqnarray}
The order of $\mathbf C$ and $\boldsymbol{\bar \Phi}_{k,i}$ are exchangeable since both are diagonal, we can thus obtain the following expression
\begin{eqnarray}\label{eq:32}
\mathbf  {\bar u}_{R,m} =  \sum_{i=0}^{M-1}{{\mathbf{\bar \Phi}}_{k,m}^H\mathcal{F}_k\bar{\bar{\mathbf G}}_{k,m,i} \mathcal{F}_k^H \mathbf{\bar \Phi}_{k,i}}\mathbf C \mathbf {\bar s}_i
\end{eqnarray}
Similarly using the same method we can derive the expression for $\mathbf{\bar{u}}_{I,m}$ as
\begin{eqnarray}\label{eq:33}
\mathbf  {\bar u}_{I,m} =  \sum_{i=0}^{M-1}{{\mathbf{\bar \Phi}}_{k,m}^H\mathcal{F}_k\bar{\tilde{\mathbf G}}_{k,m,i} \mathcal{F}_k^H \mathbf{\tilde\Phi}_{k,i}}\mathbf C \mathbf {\tilde s}_i
\end{eqnarray}
Substituting (\ref{eq:32}) and (\ref{eq:33}) into (\ref{eq:28}) yields
\begin{eqnarray}\label{eq:34}
\mathbf  {\bar y}_m \!\!\!\!\!&=& \!\!\!\!\!{\mathbf{\bar \Phi}}_{k,m}^H\mathcal{F}_k\sum_{i=0}^{M-1}{\bar{\bar{\mathbf G}}_{k,m,i} \mathcal{F}_k^H \mathbf{\bar \Phi}_{k,i}}\mathbf C \mathbf {\bar s}_i + {\mathbf{\bar \Phi}}_{k,m}^H\mathcal{F}_k\nonumber \\
&&\!\!\!\sum_{i=0}^{M-1}{\!\!\!\bar{\tilde{\mathbf G}}_{k,m,i} \mathcal{F}_k^H \mathbf{ \tilde\Phi}_{k,i}}\mathbf C \mathbf {\tilde s}_i\!+\!{\mathbf{\bar \Phi}}_{k,m}^H\mathbf {\mathcal{F}}_k\mathbf{\bar{P}}^{H}_{k,m}\mathbf{n}
\end{eqnarray}
We can further reduce (\ref{eq:34}) as
\begin{eqnarray}\label{eq:35}
\mathbf  {\bar y}_m \!\!\!\!\!&=&\!\!\!\!\! {\mathbf{\bar \Phi}}_{k,m}^H\mathcal{F}_k{\bar{\bar{\mathbf G}}_{k,m,m} \mathcal{F}_k^H \mathbf{ \bar \Phi}_{k,m}}\mathbf C \mathbf {\bar s}_m\nonumber \\
&&+\sum_{i=0, i\neq m}^{M-1}{\mathbf{\bar \Phi}}_{k,m}^H\mathcal{F}_k{\bar{\bar{\mathbf G}}_{k,m,i} \mathcal{F}_k^H \mathbf{\bar \Phi}_{k,i}}\mathbf C \mathbf {\bar s}_i\nonumber \\
&&+ \sum_{i=0}^{M-1}{\mathbf{\bar \Phi}}_{k,m}^H\mathcal{F}_k\bar{\tilde{\mathbf G}}_{k,m,i} \mathcal{F}_k^H \mathbf{ \tilde\Phi}_{k,i}\mathbf C \mathbf {\tilde s}_i\nonumber \\
&&+ {\mathbf{\bar \Phi}}_{k,m}^H\mathbf {\mathcal{F}}_k\mathbf{\bar{P}}^{H}_{k,m}\mathbf{n}
\end{eqnarray}
\subsubsection{\textit{Channel Equalization}}
We represent one tap channel equalizer as a block diagonal matrix $\mathbf E$ and is applied to the real branch signal $\mathbf {\bar{y}}_m$ as
\begin{eqnarray}\label{eq:36}
\mathbf {\bar u}_m = \mathbf E\mathbf {\bar y}_m
\end{eqnarray}
Substituting (\ref{eq:35}) into (\ref{eq:36}) we get the equalized signal $\mathbf {\bar{u}}_m$ as
\begin{eqnarray}\label{eq:37}
\mathbf  {\bar u}_m \!\!\!\!\!&=&\!\!\!\!\! \mathbf E\Big({\mathbf{\bar \Phi}}_{k,m}^H\mathcal{F}_k{\bar{\bar{\mathbf G}}_{k,m,m} \mathcal{F}_k^H \mathbf{ \bar \Phi}_{k,m}}\mathbf C \mathbf {\bar s}_m\nonumber \\
&&+\sum_{i=0, i\neq m}^{M-1}{\mathbf{\bar \Phi}}_{k,m}^H\mathcal{F}_k{\bar{\bar{\mathbf G}}_{k,m,i} \mathcal{F}_k^H \mathbf{\bar \Phi}_{k,i}}\mathbf C \mathbf {\bar s}_i \nonumber \\
&&+ \sum_{i=0}^{M-1}{\mathbf{\bar \Phi}}_{k,m}^H\mathcal{F}_k{\bar{\tilde{\mathbf G}}_{k,m,i} \mathcal{F}_k^H \mathbf{ \tilde\Phi}_{k,i}}\mathbf C \mathbf {\tilde s}_i\Big)\nonumber \\
&&+\mathbf E{\mathbf{\bar \Phi}}_{k,m}^H\mathbf {\mathcal{F}}_k\mathbf{\bar{P}}^{H}_{k,m}\mathbf{n}
\end{eqnarray}
Eq (\ref{eq:37}) can be written as
\begin{eqnarray}\label{eq:38}
\mathbf  {\bar u}_m \!\!\!\!\!&=&\!\!\!\!\! \mathbf E\mathbf C\Big({\mathbf{\bar \Phi}}_{k,m}^H\mathcal{F}_k{\bar{\bar{\mathbf G}}_{k,m,m} \mathcal{F}_k^H \mathbf{ \bar \Phi}_{k,m}}\mathbf {\bar s}_m\nonumber \\
&&+\sum_{i=0, i\neq m}^{M-1}{\mathbf{\bar \Phi}}_{k,m}^H\mathcal{F}_k{\bar{\bar{\mathbf G}}_{k,m,i} \mathcal{F}_k^H \mathbf{\bar \Phi}_{k,i}} \mathbf {\bar s}_i \nonumber \\
&&+ \sum_{i=0}^{M-1}{\mathbf{\bar \Phi}}_{k,m}^H\mathcal{F}_k{\bar{\tilde{\mathbf G}}_{k,m,i} \mathcal{F}_k^H \mathbf{ \tilde\Phi}_{k,i}}\mathbf {\tilde s}_i\Big)\nonumber \\
&&+\mathbf E{\mathbf{\bar \Phi}}_{k,m}^H\mathbf {\mathcal{F}}_k\mathbf{\bar{P}}^{H}_{k,m}\mathbf{n}
\end{eqnarray}
where $\mathbf E$ can be either ZF or MMSE based linear channel equalizer \cite{1468479}
\begin{eqnarray}\label{eq:39}
\mathbf {E} = {\mathbf C}^H(\mathbf C \mathbf C^H + \nu\sigma^2/\delta^2\mathbf{I})^{-1}
\end{eqnarray}
where $\nu = 0$ for ZF while $\nu = 1$ is for MMSE case.\\
With a simple ZF equalization i.e. $\mathbf E = (\mathbf C)^{-1}$, we can write (\ref{eq:38}) as
\begin{eqnarray}\label{eq:40}
\mathbf  {\bar u}_m \!\!\!\!\!&=&\!\!\!\!\! {\mathbf{\bar \Phi}}_{k,m}^H\mathcal{F}_k{\bar{\bar{\mathbf G}}_{k,m,m} \mathcal{F}_k^H \mathbf{ \bar\Phi}_{k,m}}\mathbf {\bar s}_m\nonumber \\
&&+\sum_{i=0, i\neq m}^{M-1}\underbrace{{\mathbf{\bar \Phi}}_{k,m}^H\mathcal{F}_k{\bar{\bar{\mathbf G}}_{k,m,i} \mathcal{F}_k^H \mathbf{\bar \Phi}_{k,i}}}_{\bar{\bar{\mathbf Q}}_{k,m,i}}\mathbf {\bar s}_i \nonumber \\
&&+\sum_{i=0}^{M-1}\underbrace{{\mathbf{\bar \Phi}}_{k,m}^H\mathcal{F}_k{\bar{\tilde{\mathbf G}}_{k,m,i} \mathcal{F}_k^H \mathbf{ \tilde\Phi}_{k,i}}}_{\bar{\tilde{\mathbf Q}}_{k,m,i}}\mathbf {\tilde s}_i\nonumber \\
&&+ \underbrace{\mathbf E{\mathbf{\bar \Phi}}_{k,m}^H\mathbf {\mathcal{F}}_k\mathbf{\bar{P}}^{H}_{k,m}\mathbf{n}}_{\mathbf{\bar u}_{noise,m}}
\end{eqnarray}
\begin{figure*}[t]
	\centering
	\begin{subfigure}[b]{0.275\textwidth}
		\includegraphics[width=\textwidth]{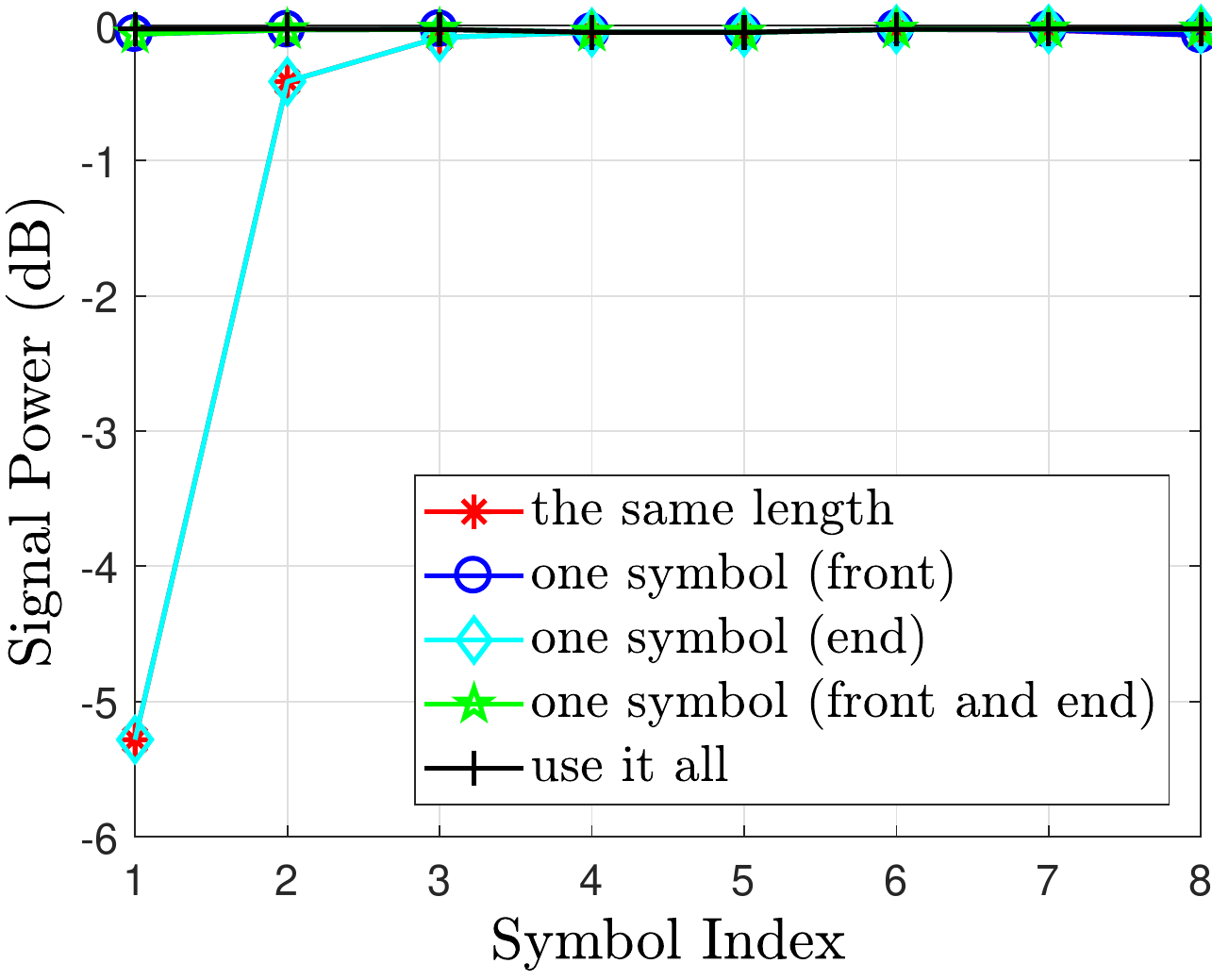}
		\caption{Signal power in $\mathcal{I}$ branch}\label{2a}
	\end{subfigure}
	~ 
	\begin{subfigure}[b]{0.275\textwidth}
		\includegraphics[width=\textwidth]{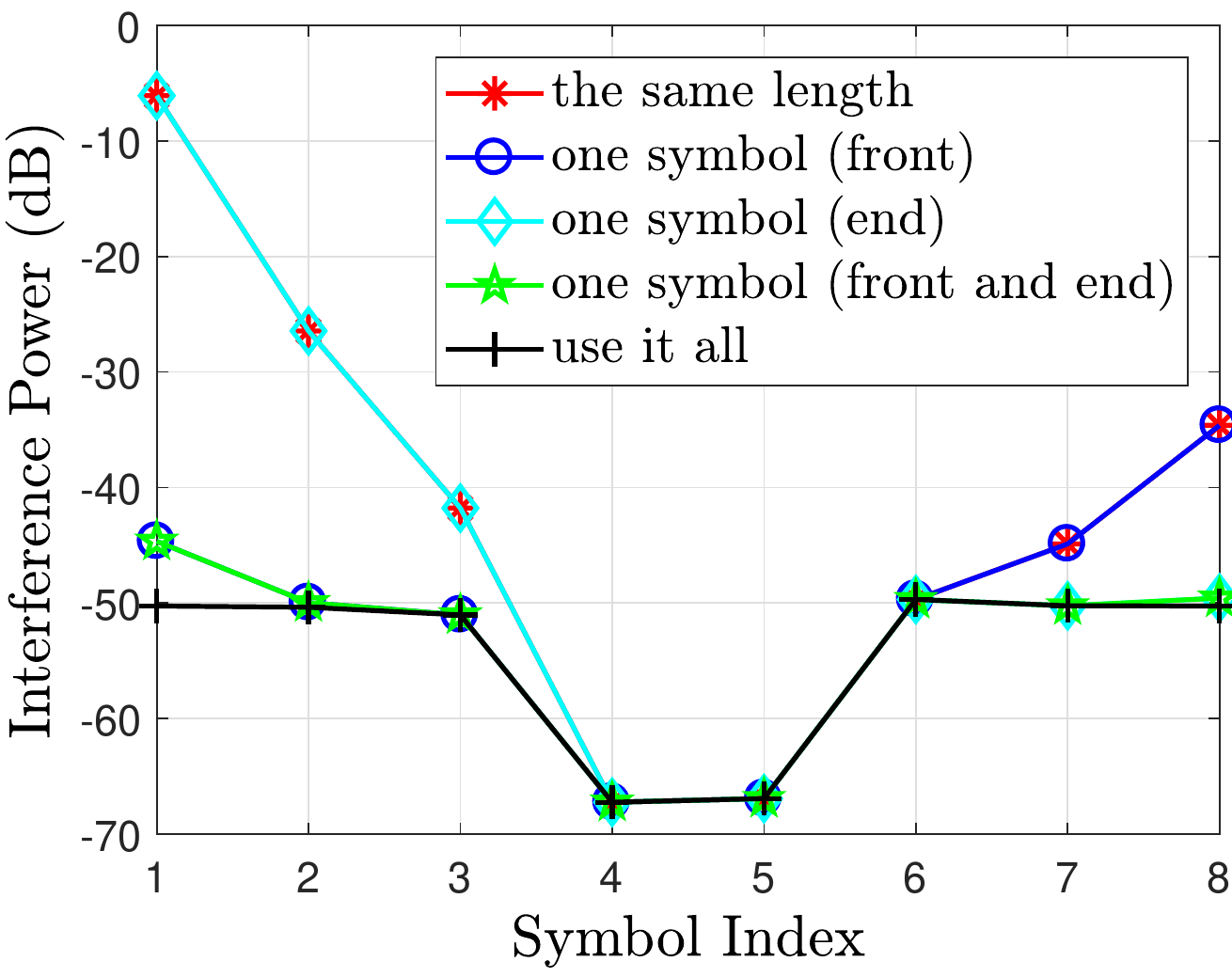}
		\caption{Interference power in $\mathcal{I}$ branch}\label{2b}
	\end{subfigure}
	~ 
	\begin{subfigure}[b]{0.275\textwidth}
		\includegraphics[width=\textwidth]{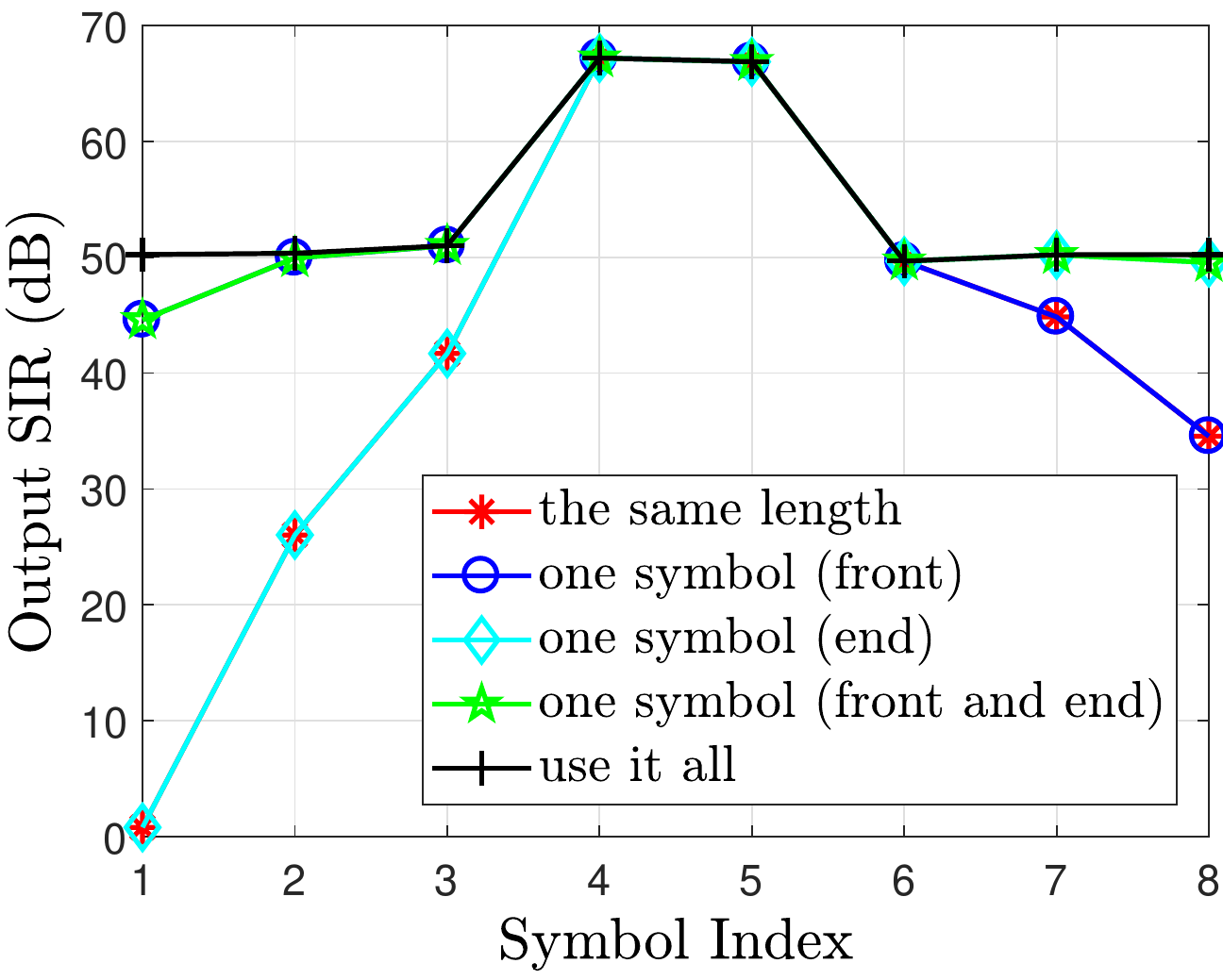}
		\caption{Output SIR in $\mathcal{I}$ branch}\label{2c}
	\end{subfigure}
	~ 
	
	\begin{subfigure}[b]{0.275\textwidth}
		\includegraphics[width=\textwidth]{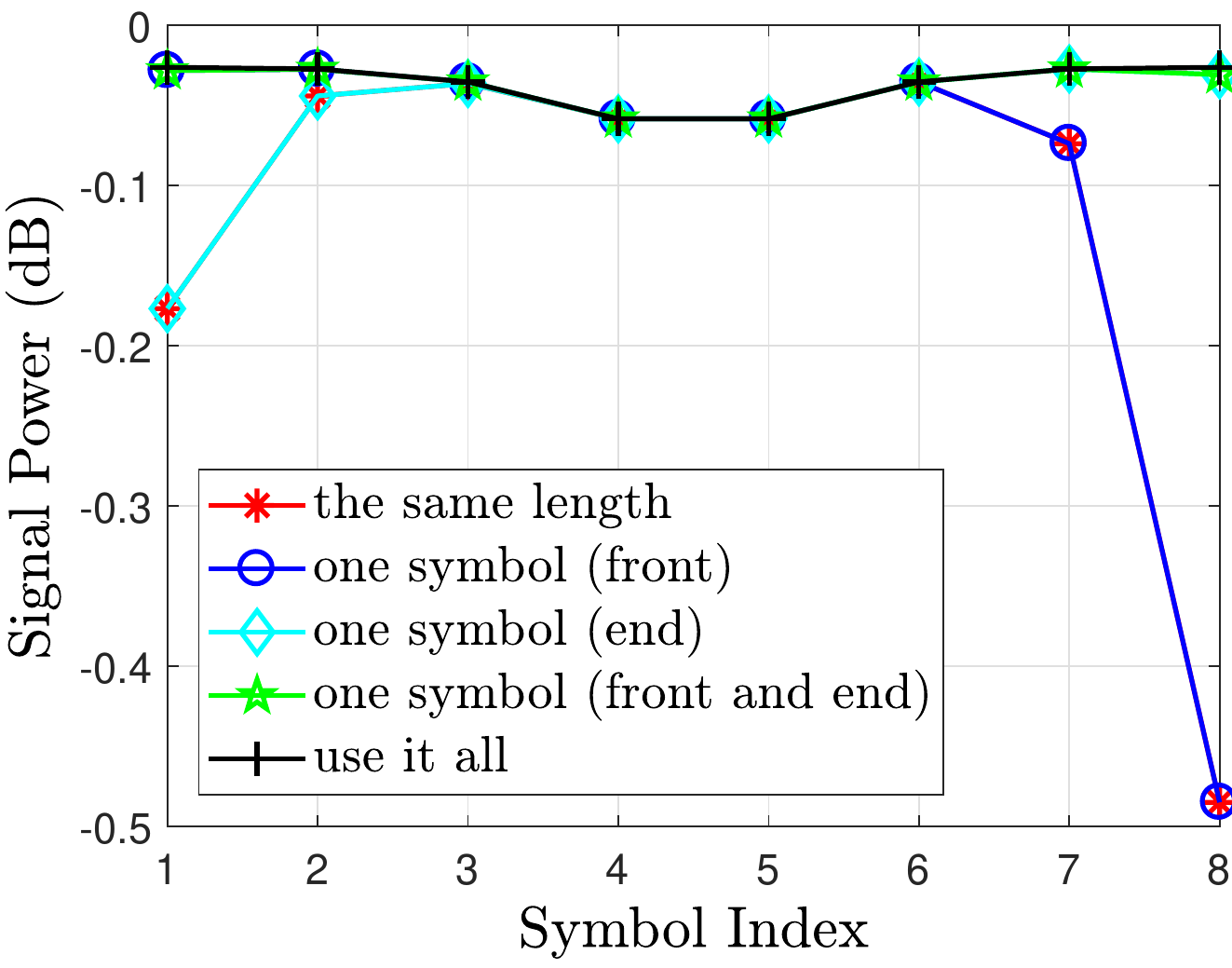}
		\caption{Signal power in $\mathcal{Q}$ branch}\label{2d}
	\end{subfigure}
	~ 
	\begin{subfigure}[b]{0.275\textwidth}
		\includegraphics[width=\textwidth]{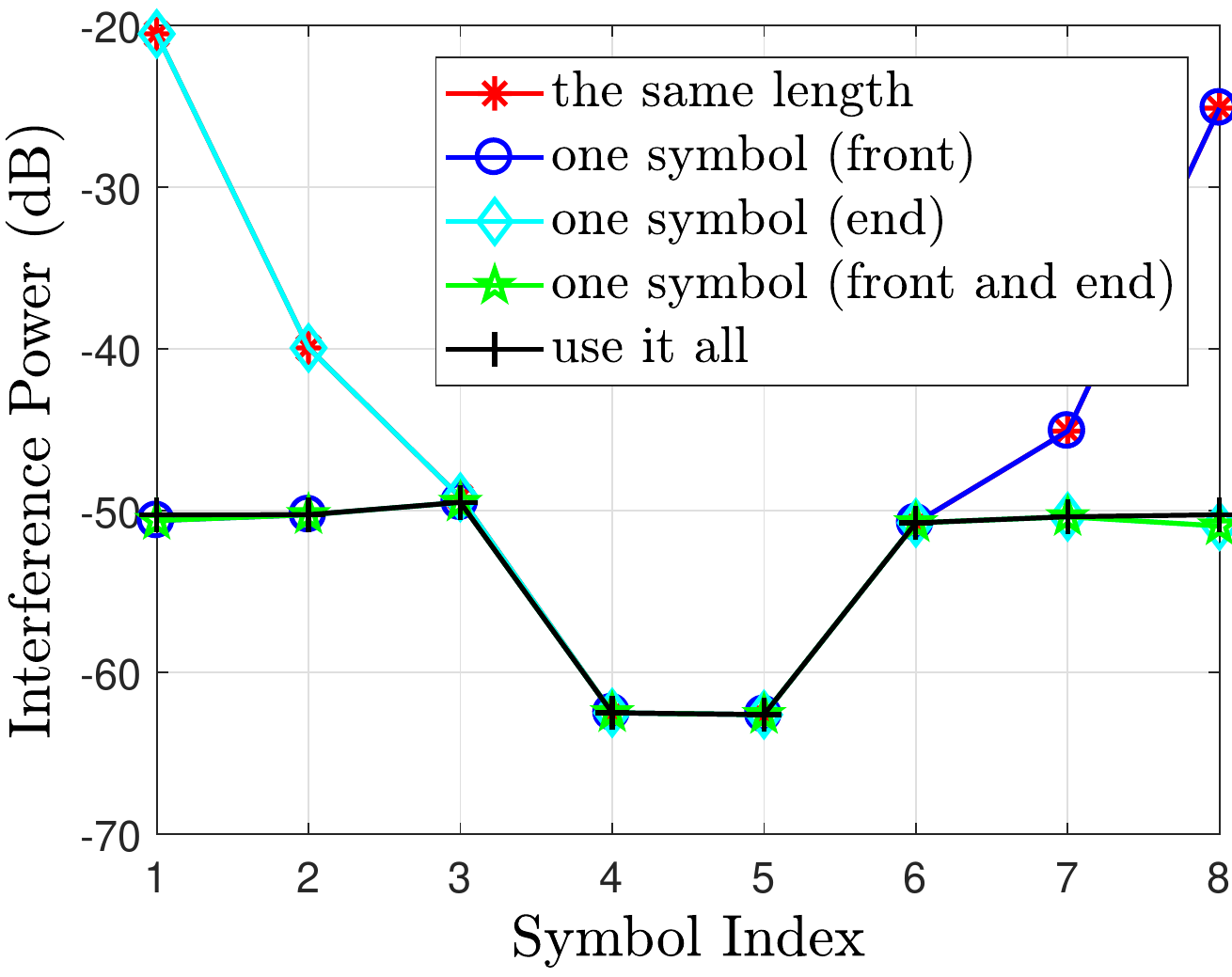}
		\caption{Interference power in $\mathcal{Q}$ branch}\label{2e}
	\end{subfigure}
	~ 
	\begin{subfigure}[b]{0.275\textwidth}
		\includegraphics[width=\textwidth]{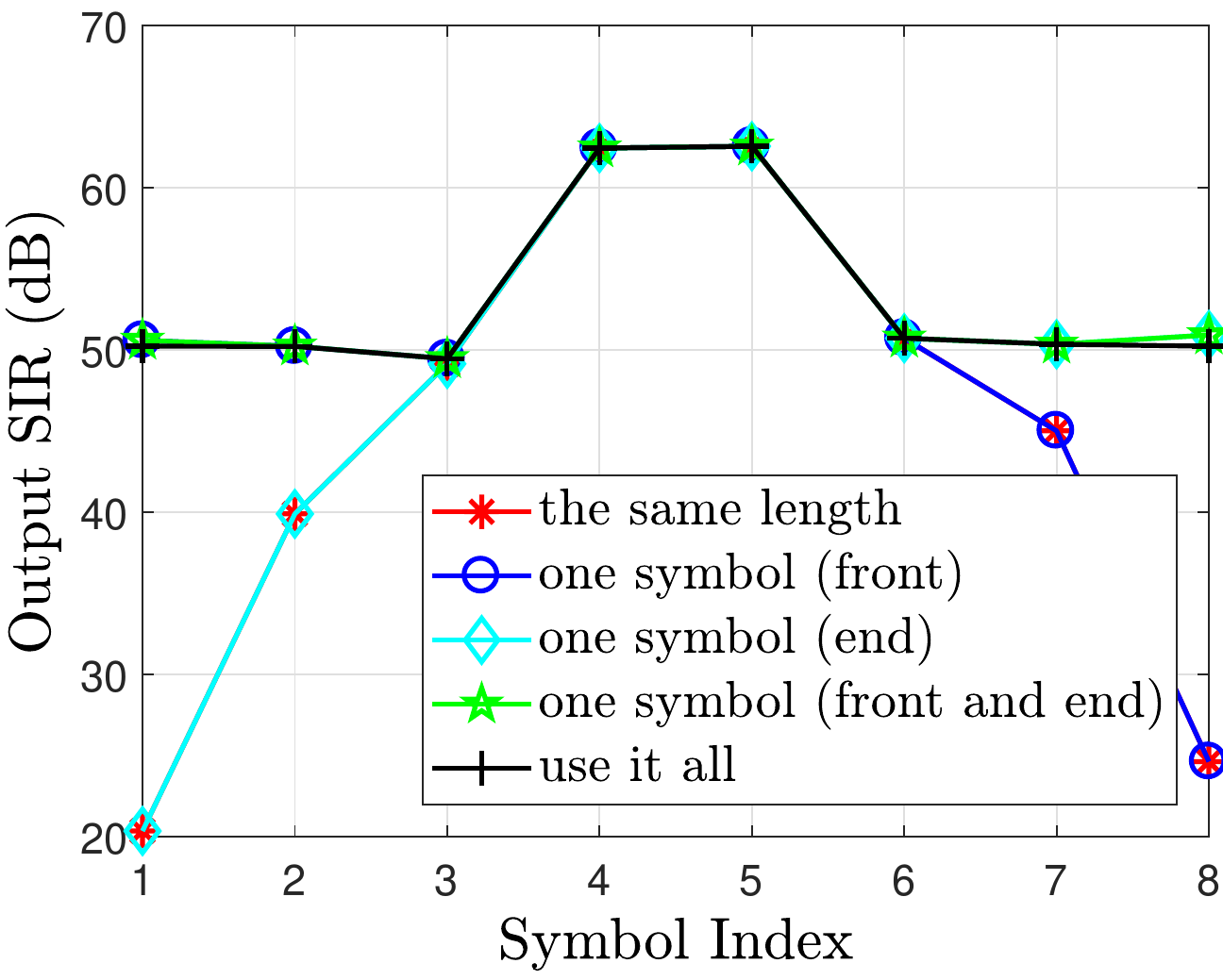}
		\caption{Output SIR in $\mathcal{Q}$ branch}\label{2f}
	\end{subfigure}
	\caption{{Signal and interference power with output SIR in real and imaginary branches $(\!K\!=\!6\!)$}}\label{fig:fig2}
\end{figure*}
\section{Finite Filter Length and Filter Output Truncation Analysis}\label{sec4}
This section presents the impact of finite filter length and FOT on the system performance. We will first consider the case with infinite filter length with no FOT and then we will extend our findings to derive the interferences caused by truncating the infinite filter length.
\subsection{\textit{Infinite Filter Length $(K=\infty)$ with no FOT}} 
In this case the autocorrelation and cross correlation matrices used in (\ref{eq:40}) can now be written as $\bar {\bar{\mathbf G}}_k = \mathbf{\bar{P}}_{k,orig}^{H}\mathbf{\bar{P}}_{k,orig}$ and $\bar {\tilde{\mathbf G}}_k = \mathbf{\bar{P}}_{k,orig}^{H}\mathbf{\tilde{P}}_{k,orig}$ respectively. According to the orthogonality of FBMC with infinite filter length \cite{7549072}, $\bar{\bar{\mathbf Q}}_{k,m,i}$ and $\bar{\tilde{\mathbf Q}}_{k,m,i}$ defined in (\ref{eq:40}) have the following property:
 \begin{eqnarray}\label{eq:41}
\bar{\bar{\mathbf Q}}_{k,m,i} \!\!\!\!\!&=&\!\!\!\!\! \left\{\!\!\!
            \begin{array}{lcl}
             \mathbf {I} + j\Im\{{\bar{\bar{\mathbf Q}}_{k,m,i}}\}  \quad \textrm{for} \quad  i=m\\
             j\Im\{{\bar{\bar{\mathbf Q}}_{k,m,i}}\}  \!\!\quad \quad \quad \textrm{for} \quad i \neq m
             \end{array}
        \right.\nonumber \\
\bar{\tilde{\mathbf Q}}_{k,m,i} \!\!\!\!\!&=&\!\!\!\!\!  j\Im\{{\bar{\tilde{\mathbf Q}}_{k,m,i}}\} \!\!\!\!\!\! \quad \quad \textrm{for} \quad i = 0,\!\cdots\!,M\!\!-\!\!1
\end{eqnarray}
Using the property of infinite filter length given in (\ref{eq:41}), we can write (\ref{eq:40}) as
\begin{eqnarray}\label{eq:42}
\mathbf  {\bar u}_m \!\!\!\!\!&=&\!\!\!\!\! \mathbf {\bar s}_m+\underbrace{j\Big[\sum_{i=0}^{M-1}\!\!\Im\{{\bar{\bar{\mathbf Q}}_{k,m,i}}\}\mathbf {\bar s}_i\!\!+\!\!\!\sum_{i=0}^{M-1}\!\!\!\Im\{{\bar{\tilde{\mathbf Q}}_{k,m,i}}\}\mathbf {\tilde s}_i\Big]}_{\mathbf {\bar u}_{intri,m}}\nonumber \\
&&+ \mathbf{\bar u}_{noise,m}
\end{eqnarray}
\indent The $\mathbf {\bar u}_{intri,m}$ term is the pure imaginary \textit{intrinsic interference} that is inherent in the FBMC system. This interference can be avoided by taking the real part of (\ref{eq:42}). Hence, we can write (\ref{eq:42}) as
\begin{eqnarray}\label{eq:43}
\Re\{\mathbf  {\bar u}_m\} \!\!\!\!&=&\!\!\!\! \mathbf {\bar s}_m+\Re\{\mathbf{\bar u}_{noise,m}\}
\end{eqnarray}
Eq (\ref{eq:43}) shows that with infinite filter length and no truncation, the actual transmitted symbol i.e. $\mathbf {\bar s}_m$ can be recovered without any ISI or ICI. The term $\Re(\mathbf{\bar u}_{noise,m})$ is the real part of the processed noise. If we take the real part of (\ref{eq:41}), the property is then simplified as
\begin{eqnarray}\label{eq:44}
\Re\{\bar{\bar{\mathbf Q}}_{k,m,i}\} \!\!\!\!\!&=&\!\!\!\!\! \left\{
            \begin{array}{lcl}
             \mathbf {I} \quad \textrm{for} \quad  i=m\\
             \mathbf 0  \quad \textrm{for} \quad i \neq m
             \end{array}
        \right.\nonumber \\
\Re\{\bar{\tilde{\mathbf Q}}_{k,m,i}\} \!\!\!\!\!&=&\!\!\!\!\!\mathbf 0 \!\!\!\! \quad \quad \quad \textrm{for} \quad i = 0,\cdots,M\!\!-\!\!1
 \end{eqnarray}
The simplified property satisfies the result obtain in (\ref{eq:43}). 
\subsection{\textit{Finite Filter Length $(K\!\ne\!\infty)$ with FOT}} As it is impractical to use infinite filter length from implementation point of view, we now consider the practical case where we consider a finite filter length $(K\!\ne\!\infty)$ with FOT. In this case the autocorrelation and cross correlation matrices given in (\ref{eq:40}) are now defined using the truncated matrices define in (\ref{eq:3}) i.e. $\bar {\bar{\mathbf G}}_k = \mathbf{\bar{P}}_k^{H}\mathbf{\bar{P}}_k$ and $\bar {\tilde{\mathbf G}}_k = \mathbf{\bar{P}}_k^{H}\mathbf{\tilde{P}}_k$ respectively. In this case, (\ref{eq:44}) will now be modified as
\begin{eqnarray}\label{eq:45}
\Re\{\bar{\bar{\mathbf Q}}_{k,m,i}\} \!\!\!\!&=&\!\!\!\!\! \left\{
            \begin{array}{lcl}
             \mathbf {I}+\Re\{\Delta\bar{\bar{\mathbf Q}}_{k,m,m}\} \quad \textrm{for} \quad  i=m\\
             \Re\{\Delta\bar{\bar{\mathbf Q}}_{k,m,i}\}  \quad \quad\quad \textrm{for} \quad i \neq m
             \end{array}
        \right.\nonumber \\
\Re\{\bar{\tilde{\mathbf Q}}_{k,m,i}\} \!\!\!\!\!&=&\!\!\!\!\!\Re\{\Delta\bar{\tilde{\mathbf Q}}_{k,m,i}\} \!\! \quad \textrm{for} \quad i = 0,\!\!\cdots\!\!,M\!\!-\!\!1
\end{eqnarray}
where $\Delta{\bar{\bar{\mathbf Q}}_{k,m,i}}={\mathbf{\bar \Phi}}_{k,m}^H\mathcal{F}_k{\Delta\bar{\bar{\mathbf G}}_{k,m,i} \mathcal{F}_k^H \mathbf{\bar \Phi}_{k,i}}$ and $\Delta{\bar{\tilde{\mathbf Q}}_{k,m,i}}={\mathbf{\bar \Phi}}_{k,m}^H\mathcal{F}_k{\Delta\bar{\tilde{\mathbf G}}_{k,m,i} \mathcal{F}_k^H \mathbf{\tilde \Phi}_{k,i}}$ in which $\Delta\bar{\bar{\mathbf G}}_{k,m,i}$ and $\Delta\bar{\tilde{\mathbf G}}_{k,m,i}$ are the error matrices due to the finite filter length and truncating effect. 
Hence, Eq (\ref{eq:43}) will now be modified as
\begin{eqnarray}\label{eq:46}
\Re\{\mathbf  {\bar u}_m\} \!\!\!\!\!&=&\!\!\!\!\! \mathbf {\bar s}_m+\sum_{i=0}^{M-1}\Re\{\Delta{\bar{\bar{\mathbf Q}}_{k,m,i}}\}\mathbf {\bar s}_i\nonumber \\
&&\!\!\!\!\!+\sum_{i=0}^{M-1}\!\!\!\Re\{\Delta{\bar{\tilde{\mathbf Q}}_{k,m,i}}\}\mathbf {\tilde s}_i\!+\!\Re\{\mathbf{\bar u}_{noise,m}\!\}
\end{eqnarray}
\indent The variance of elements in the error matrices not only depends on the filter length $K$ and the truncation number $i_F$ and $i_R$, but more importantly on the odd or even value of $K$. The truncation causes the filter correlation matrices to be unsaturated at both the edges i.e. the symbols at the start and at the end of the block will experience truncation effect while the truncation causes the filter correlation matrix to be saturated in the middle part. Hence, the symbols in the middle of the filtered MIMO-FBMC block are least effected. This can be confirmed from \cite{7390479}, where we have demonstrated that with finite filter length $(K\!=\!6)$, the filter output contains $K\!-\!1$ symbols and that these extra tails at the edges of the FBMC block have small average energy compared to the middle part of the block.
\subsection{\textit{Filter Output Truncation (FOT) Analysis}}
To analyze the impact of these factors on the filter output truncation, we consider the following cases. We first consider the even value of filter length $(K=6)$, which will introduce $K-1$ tails i.e. 5 extra symbols at the output of the transmit filter. Also we have assumed $M=8$ i.e. symbols per block at the input of the filter. \textit{Note that this value of $M$ is considered just as an example and does not affect the outcomes of the analysis}.\vspace*{.1cm}
\begin {enumerate}[a)]
\item {\bf{Use it all:}} No cut at all ($i_F\!=\!0$ , $i_R\!=\!0$), i.e., input 8 symbols and output 13 symbols.
\item {\bf{One symbol (front and end):}} Cut 2 at the front and 1 at the end ($i_F\!=\!2$ , $i_R\!=\!1$), i.e., input 8 symbols and output 10 symbols.
\item {\bf{One symbol (front):}} Cut 2 at the front and 2 at the end ($i_F\!=\!2$ , $i_R\!=\!2$), i.e., input 8 symbols and output 9 symbols.
\item {\bf{One symbol (end):}} Cut 3 at the front and 1 at the end ($i_F\!=\!3$ , $i_R\!=\!1$), i.e., input 8 symbols and output 9 symbols.
\item {\bf{The same length:}} Cut the front 3 and last 2 symbols ($i_F\!=\!3$ , $i_R\!=\!2$), to keep the number of symbols the same i.e. input 8 symbols and output 8 symbols.
\end{enumerate}
Fig. \ref{fig:fig2} shows the desired signal and interference powers for real and imaginary branches in case of finite filter length $(\!K\!=\!6)$ with different FOT scenarios. The observations drawn from Fig. \ref{fig:fig2} regarding the aforementioned cases are discussed as follows
\begin{itemize}
\item \textit{Use it all} case i.e. no truncation can achieve very good performance for both real or imaginary branches. As in this case, the second and third terms in (\ref{eq:46}) will not exist and therefore the desired symbols are free from interference terms.
\item \textit{One symbol (front and end)} case can achieve similar performance as in \textit{use it all} case, only marginal difference is at the edge symbols. This is because the one symbol at the front has significant energy as compared to the other tails \cite{7390479}. Leaving this symbol at the front will significantly reduce the interference level and the effect of cutting other two symbols at the front and one at the end has much less affect on the neighboring symbols as can be seen from Fig. {\ref{2b}}.
\item \textit{One symbol (front)} case introduces interference at the last symbols i.e. $m=7,8$ compared to the \textit{one symbol (front and end)} case. This loss is tolerable as the signal power loss and the increase in the interference level for $m=7$ and $m=8$ are insignificant as can be seen from  Fig. \ref{2b}. These losses are acceptable as we are avoiding an extra symbol overhead compared to the $\textit{one symbol (front and end)}$ case. This performance loss at the last symbols is due to the truncation at the end of the filter that introduces interference in the last two symbols.
\item However, \textit{One symbol (end)} case does not work as the signal power for $m=1$ is reduced and the interference level has increased significantly which are both unacceptable. It is because in this case we are truncating the front part of the filter that discards all the symbols at the front of the block and introduces significant interference in the neighboring symbols. Hence, leaving one symbol at the end is not a good strategy.
\item In \textit{the same length} case, the desired signal power and interference power for the symbols at the edges $(m=1,2$ and $7,8)$ are affected significantly. This is because the extra symbols at the start and the end of the block are truncated that affects their neighboring symbols. In this case, the second and third terms in (\ref{eq:46}) will exist and as a result, the detected symbols will be effected by these interference terms.
\end{itemize}
\begin{figure*}[t]
	\centering
	\begin{subfigure}[b]{0.275\textwidth}
		\includegraphics[width=\textwidth]{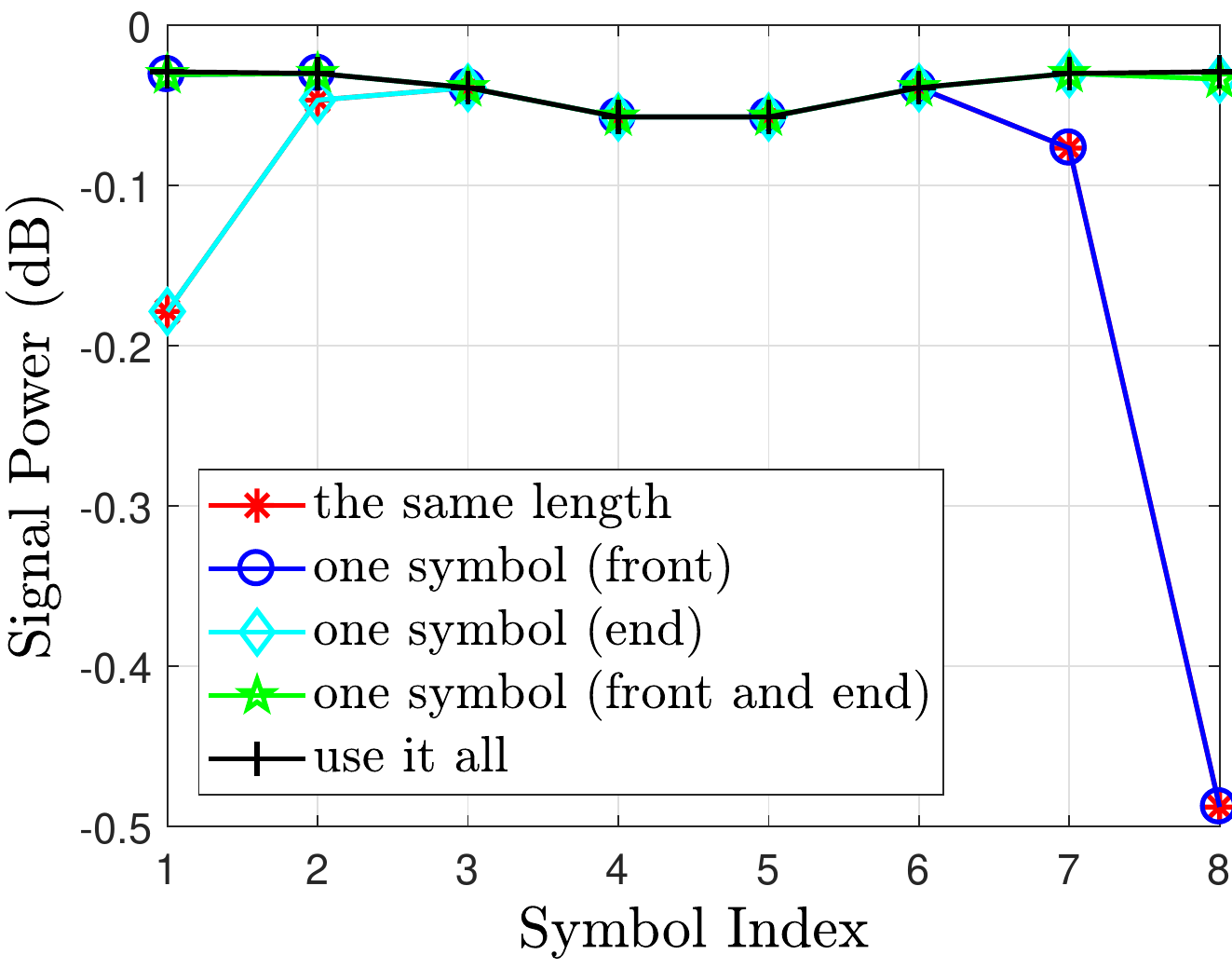}
		\caption{Signal power in $\mathcal{I}$ branch}\label{3a}
	\end{subfigure}
	~ 
	\begin{subfigure}[b]{0.275\textwidth}
		\includegraphics[width=\textwidth]{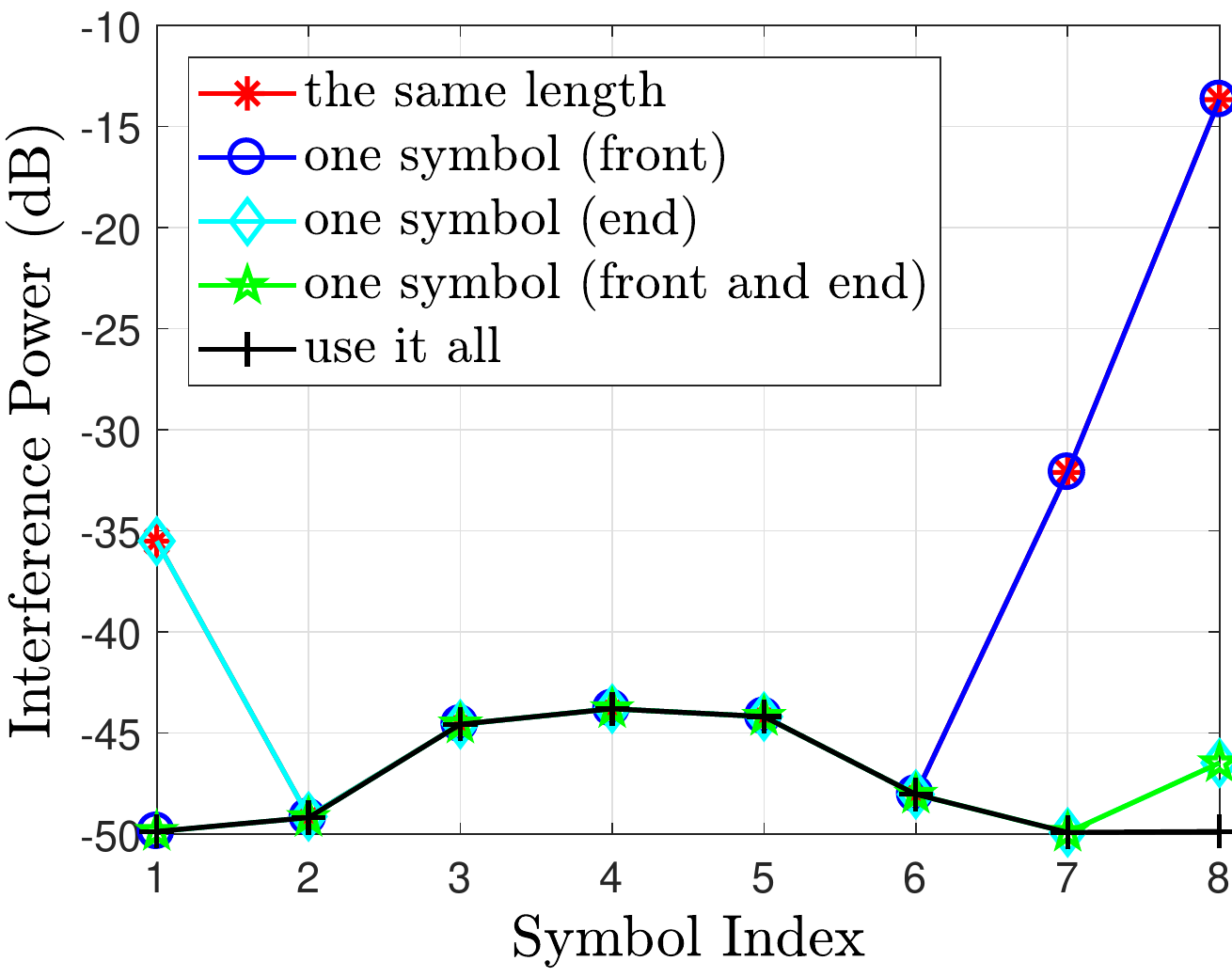}
		\caption{Interference power in $\mathcal{I}$ branch}\label{3b}
	\end{subfigure}
	~ 
	\begin{subfigure}[b]{0.275\textwidth}
		\includegraphics[width=\textwidth]{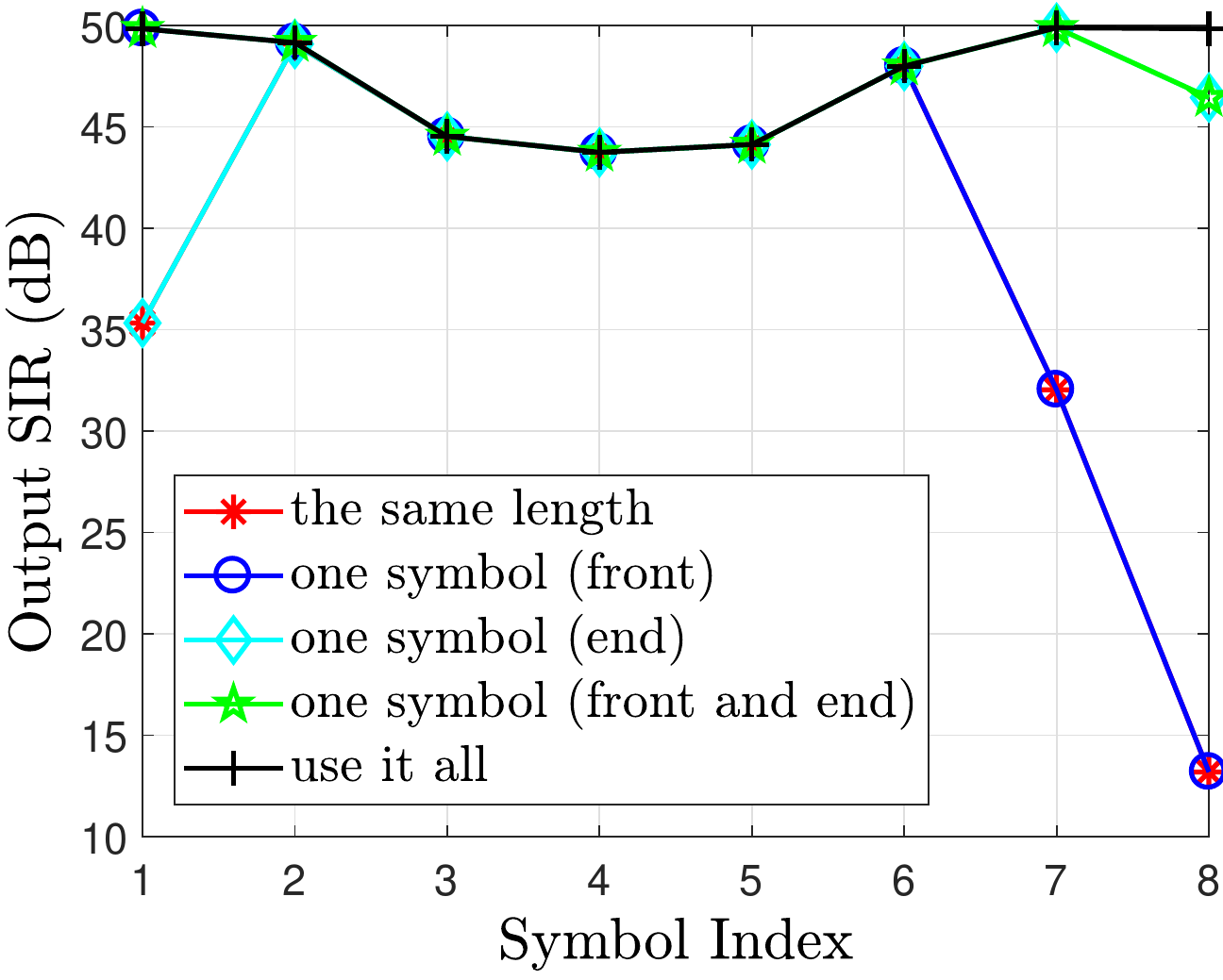}
		\caption{Output SIR in $\mathcal{I}$ branch}\label{3c}
	\end{subfigure}
	~ 
	
	\begin{subfigure}[b]{0.275\textwidth}
		\includegraphics[width=\textwidth]{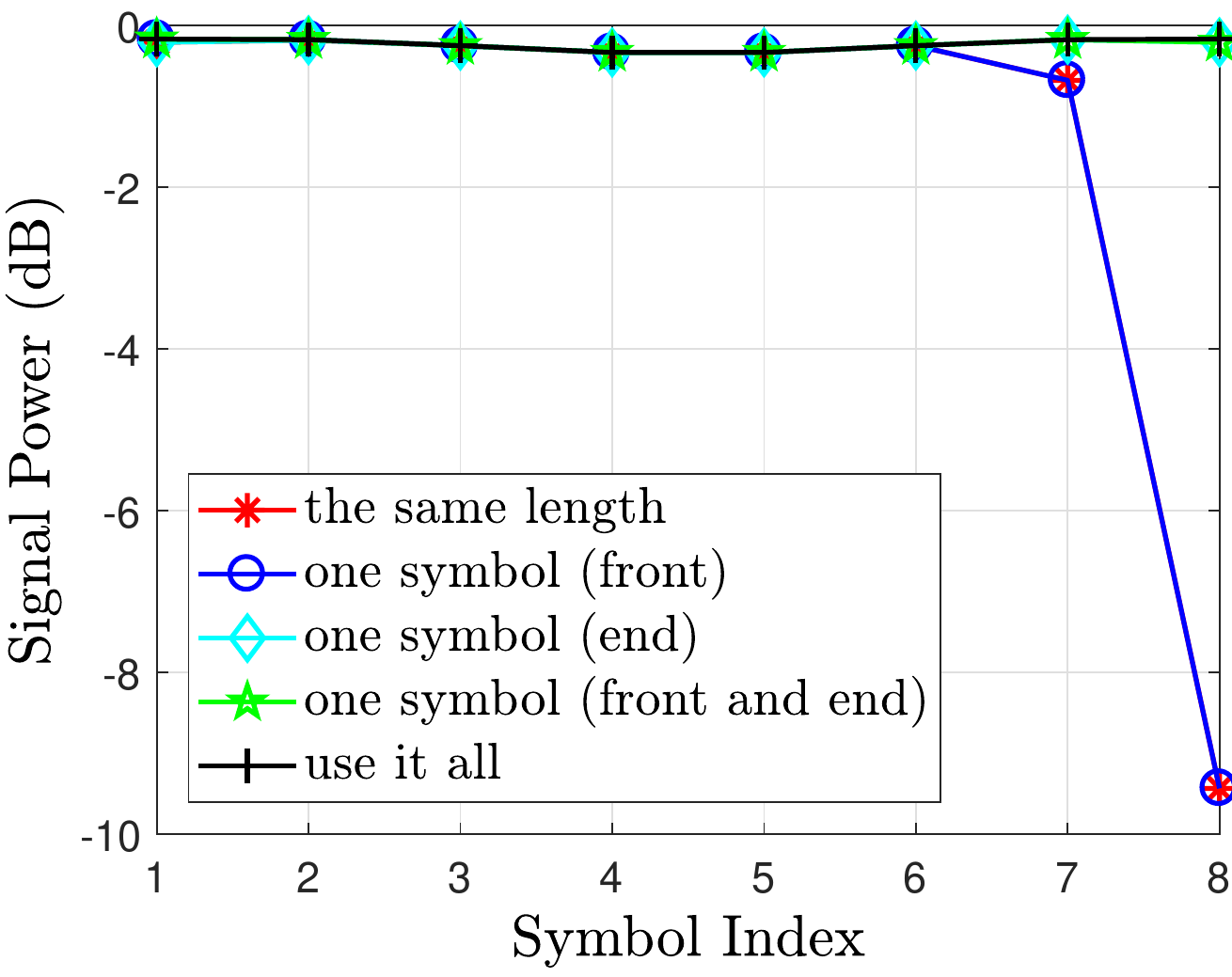}
		\caption{Signal power in $\mathcal{Q}$ branch}\label{3d}
	\end{subfigure}
	~ 
	\begin{subfigure}[b]{0.275\textwidth}
		\includegraphics[width=\textwidth]{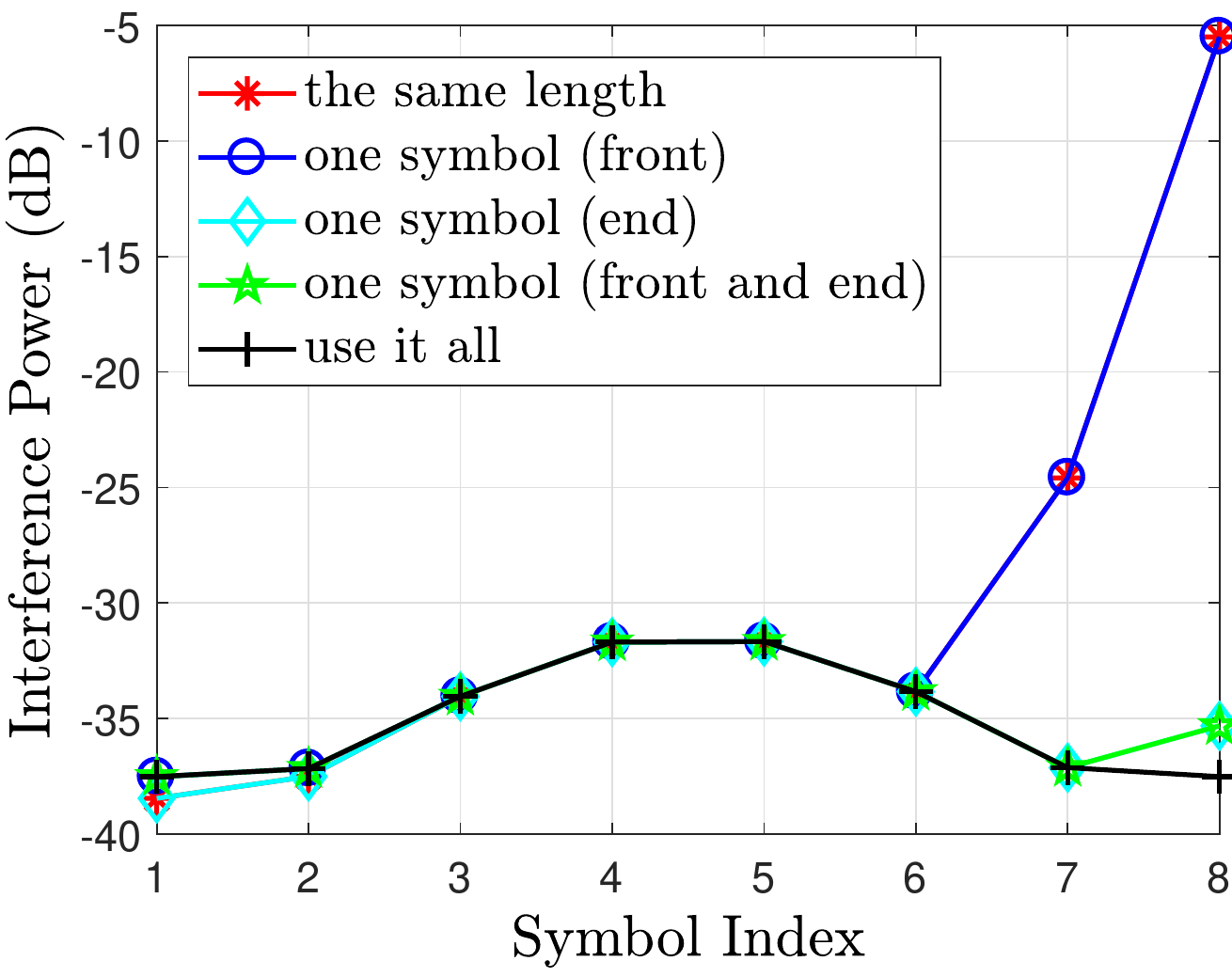}
		\caption{Interference power in $\mathcal{Q}$ branch}\label{3e}
	\end{subfigure}
	~ 
	\begin{subfigure}[b]{0.275\textwidth}
		\includegraphics[width=\textwidth]{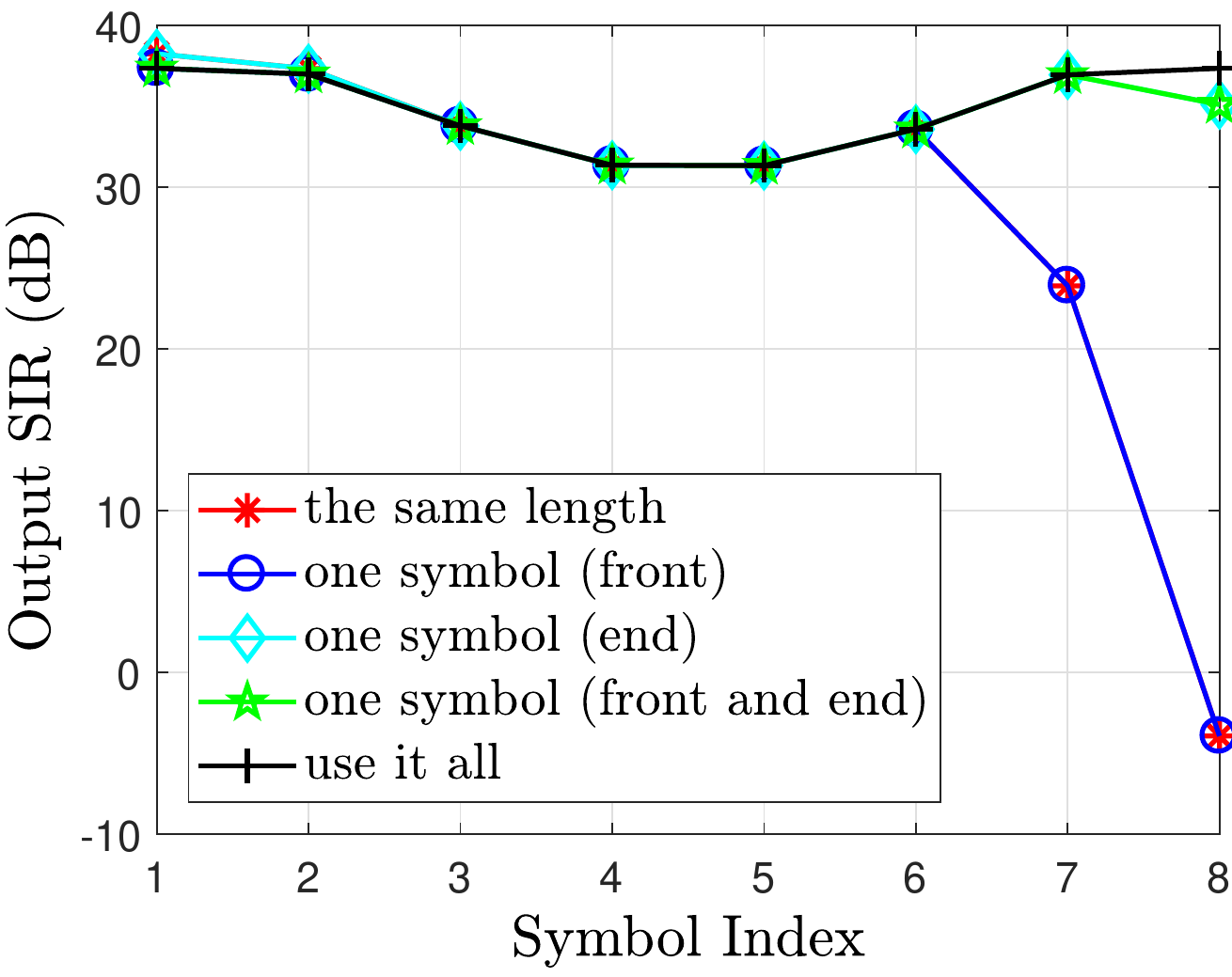}
		\caption{Output SIR in $\mathcal{Q}$ branch}\label{3f}
	\end{subfigure}
	\caption{{Signal and interference power with output SIR in real and imaginary branches $(\!K\!=\!5\!)$}}\label{fig:fig4}
\end{figure*}

\indent The output signal to interference ratio (SIR) for real and imaginary branches is illustrated in Fig. \ref{2c} and Fig. \ref{2f} respectively, where we can see that with a finite filter length $(K\!=\!6)$, the best SIR can be obtained with \textit{use it all} case; however, the overhead is quite high in this case. While \textit{the same length} case can completely remove the overhead but significantly reduces the SIR of the symbols at the edges. A good balance is to adopt the \textit{one symbol (front)} case for even $K$ which offers an acceptable trade-off between the overhead and the performance.\vspace*{.1cm}\\
\indent However, the observations are totally reversed when we consider the odd number of filter length e.g. $(K\!=\!5)$. In this case the last symbol in the imaginary branch is significantly affected by the FOT as can be seen from Fig. \ref{fig:fig4}. The \textit{one symbol (end)} case is now more effective in case of odd filter length as it provides better SIR compared to the other cases as can be seen from Fig. \ref{3c} and Fig. \ref{3f}.\\
Since the target branch and symbol are totally different for odd and even $K$, in the next section, we will focus on the even $K$ only for proposing the compensation algorithm. The compensation algorithm for the odd $K$ can be derived using the same approach.
\section{Proposed Compensation Algorithm}\label{sec5}
Although adding \textit{one symbol (front)} case can provide acceptable performance (SIR$>$20dB). However, this approach is valid only when the block size $M$ is large. For instance when $M=20$, the total overhead is only 5$\%$ and this percentage further drops when $M$ goes to larger value \cite{7390479}. However, considering different traffic models and also the latency of the data, the solution that \textit{one symbol (front)} case may cause significant overhead e.g. with moderate $M=5$, the total overhead is 20$\%$, which is very inefficient.\vspace*{.1cm}\\
\indent In order to overcome this inefficiency for moderate $M$, we propose a compensation approach which allows complete removal of the overhead caused by the filtering operation. Note that when $K$ is even, if we consider \textit{the same length} case only the first symbol on the $\mathcal{I}$ branch has unacceptable level of SIR whereas the corresponding symbol on the $\mathcal{Q}$ branch has sufficient SIR level (20dB) as can be see from Fig. \ref{2c} and Fig. \ref{2f} respectively. While in the odd $K$ case, the situation is opposite (only the last symbol on the $\mathcal{Q}$ branch has unacceptable level of SIR) as can be seen from Fig. \ref{3f}. With this observation, we can state that all of the other symbols (both real and imaginary, except the first real symbol for even $K$ or last imaginary symbol for odd $K$) can be easily detected.\vspace*{.1cm} \\
\indent Considering the even $K$ case with the assumption that the channel is known and that we only need to compensate the first symbol in the real branch to have sufficient SIR value to detect all the symbols. According to (\ref{eq:46}), the first $\mathcal{I}$ branch symbol can be written as
\begin{eqnarray}\label{eq:47}
\Re\{\mathbf  {\bar u}_0\} \!\!\!\!\!&=&\!\!\!\!\! \mathbf {\bar s}_0+ \sum_{i=0}^{M-1}\Re\{\mathbf{\bar\Phi}^H_{k,0}\mathcal{F}_k{\Delta\bar{\bar{\mathbf G}}_{k,0,i} \mathcal{F}_k^H \mathbf{ \bar\Phi}_{k,i}}\}\mathbf {\bar s}_i\nonumber \\
&&+\sum_{i=0}^{M-1}\Re\{\mathbf{\bar\Phi}^H_{k,0}\mathcal{F}_k{\Delta\bar{\tilde{\mathbf G}}_{k,0,i} \mathcal{F}_k^H \mathbf{\tilde \Phi}_{k,i}}\}\mathbf {\tilde s}_i \nonumber \\
\!\!\!\!\!&=&\!\!\!\!\! \mathbf {\bar s}_0+\Re\{\mathbf{\bar \Phi}^H_{k,0}\mathcal{F}_k{\Delta\bar{\bar{\mathbf G}}_{k,0,0} \mathcal{F}_k^H \mathbf{\bar \Phi}_{k,0}}\}\mathbf {\bar s}_0\nonumber \\
&&+\sum_{i=1}^{M-1}\Re\{\mathbf{\bar\Phi}^H_{k,0}\mathcal{F}_k{\Delta\bar{\bar{\mathbf G}}_{k,0,i} \mathcal{F}_k^H \mathbf{ \bar\Phi}_{k,i}}\}\mathbf {\bar s}_i \nonumber \\
&&+\sum_{i=0}^{M-1}\Re\{\mathbf{\bar\Phi}^H_{k,0}\mathcal{F}_k{\Delta\bar{\tilde{\mathbf G}}_{k,0,i} \mathcal{F}_k^H \mathbf{\tilde \Phi}_{k,i}}\}\mathbf {\tilde s}_i
\end{eqnarray}
\indent The first term in (\ref{eq:47}) is the desired signal, the second term is the ICI and the third and fourth terms are the ISI caused by the $\mathcal{I}$ and $\mathcal{Q}$ branches respectively. For simplicity, we omit the noise term. In order to improve the SIR of the first symbol in the $\mathcal{I}$ branch, we need to compensate the ICI and the ISI terms at the receiver. For this, we need to find the compensation matrices i.e. $\Delta\bar{\bar{\mathbf G}}_{k,0,i}$ and $\Delta\bar{\tilde{\mathbf G}}_{k,0,i}$ in (\ref{eq:47}). Note that the $\Delta\bar{\bar{\mathbf G}}_{k,0,i}$ and $\Delta\bar{\tilde{\mathbf G}}_{k,0,i}$ are caused by the FOT which brings significant SIR reduction for some symbols. To derive the matrices, we define the perfect autocorrelation matrices $\bar{\bar{\mathbf G}}_{k,orig}$ and $\bar{\tilde{\mathbf G}}_{k,orig}$ as follows
\begin{eqnarray}\label{eq:48}
\bar{\bar{\mathbf G}}_{k,orig}\!\!\!\!&=&\!\!\!\!\mathbf{\bar{P}}^H_{k,orig}\mathbf{\bar{P}}_{k,orig}=\begin{bmatrix}
{\mathbf {\bar{P}}_{k,i_F}}^H\!\!&\!\!\mathbf {\bar{P}}_k^H\!\!&\!\!{\mathbf {\bar{P}}_{k,i_R}}^H
\end{bmatrix}\!\!
\begin{bmatrix}
{\mathbf {\bar{P}}_{k,i_F}} \\
\mathbf {\bar{P}}_k \\
{\mathbf {\bar{P}}_{k,i_R}} \\
\end{bmatrix} \nonumber \\
\!\!\!\!&=&\!\!\!\!{\mathbf {\bar{P}}_{k,i_F}}^H\mathbf {\bar{P}}_{k,i_F}+\mathbf {\bar{P}}_k^H\mathbf {\bar{P}}_k+{\mathbf {\bar{P}}_{k,i_R}}^H\mathbf {\bar{P}}_{k,i_R}
\end{eqnarray}
Similarly,
\begin{eqnarray}\label{eq:49}
\bar{\tilde{\mathbf G}}_{k,orig}\!\!\!\!&=&\!\!\!\!{\mathbf {\bar{P}}_{k,i_F}}^H\mathbf {\tilde{P}}_{k,i_F}\!+\!\mathbf {\bar{P}}_k^H\mathbf {\tilde{P}}_k\!+\!{\mathbf {\bar{P}}_{k,i_R}}^H\mathbf {\tilde{P}}_{k,i_R}
\end{eqnarray}
\indent We can write the compensation matrices $\Delta\bar{\bar{\mathbf G}}_k$ and $\Delta\bar{\tilde{\mathbf G}}_k$ using the perfect autocorrelation matrices ($\bar{\bar{\mathbf G}}_{k,orig}$ and $\bar{\tilde{\mathbf G}}_{k,orig}$) and the truncated autocorrelation matrices ($\bar{\bar{\mathbf G}}_k=\mathbf{\bar{P}}_k^{H}\mathbf{\bar{P}}_k$ and $\bar {\tilde{\mathbf G}}_k= \mathbf{\bar{P}}_k^{H}\mathbf{\tilde{P}}_k$) as: 
\begin{eqnarray}\label{eq:50}
\Delta\bar{\bar{\mathbf G}}_k\!\!\!\!\!&=&\!\!\!\!\!\bar{\bar{\mathbf G}}_{k,orig}\!-\!\bar{\bar{\mathbf G}}_k\!=\!{\mathbf {\bar{P}}_{k,i_F}}^H\!\mathbf {\bar{P}}_{k,i_F}\!\!+\!\!{\mathbf {\bar{P}}_{k,i_R}}^H\!\mathbf {\bar{P}}_{k,i_R}
\end{eqnarray}
\begin{eqnarray}\label{eq:51}
\Delta\bar{\tilde{\mathbf G}}_k\!\!\!\!\!&=&\!\!\!\!\!\bar{\tilde{\mathbf G}}_{k,orig}\!-\!\bar{\tilde{\mathbf G}}_k\!=\!{\mathbf {\bar{P}}_{k,i_F}}^H\!\mathbf {\tilde{P}}_{k,i_F}\!\!+\!\!{\mathbf {\bar{P}}_{k,i_R}}^H\!\mathbf {\tilde{P}}_{k,i_R}
\end{eqnarray}
\indent Now for even $K$ case, we propose the following compensation algorithm to determine $\Delta\bar{\bar{\mathbf G}}_{k,0,i}$ and $\Delta\bar{\tilde{\mathbf G}}_{k,0,i}$ in (\ref{eq:47}) for compensating ISI in the first real symbol.
Using (\ref{eq:50}) and (\ref{eq:51}), we can determine $\Delta\bar{\bar{\mathbf G}}_{k,0,i}=\Delta\bar{\bar{\mathbf G}}_{0,i}\otimes \mathbf I_{N_{r}}$ and $\Delta\bar{\tilde{\mathbf G}}_{k,0,i}=\Delta\bar{\tilde{\mathbf G}}_{0,i}\otimes \mathbf I_{N_{r}}$ for $i=0\cdots M-1$ using (\ref{eq:3}) as
\begin{eqnarray}\label{eq:52}
\Delta\bar{\bar{\mathbf G}}_{0,0}&=&\bar {\mathbf W}_0^H\bar {\mathbf W}_0+\bar {\mathbf W}_1^H\bar {\mathbf W}_1+\bar {\mathbf W}_2^H\bar {\mathbf W}_2 \nonumber \\
\Delta\bar{\bar{\mathbf G}}_{0,1}&=&\bar {\mathbf W}_1^H\bar {\mathbf W}_0+\bar {\mathbf W}_2^H\bar {\mathbf W}_1 \nonumber \\
\Delta\bar{\bar{\mathbf G}}_{0,2}&=&\bar {\mathbf W}_2^H\bar {\mathbf W}_0 \nonumber \\
\Delta\bar{\bar{\mathbf G}}_{0,j}&=& \mathbf 0 \quad\quad\textrm{for}\quad 3\le j \le M-1
\end{eqnarray}
and
\vspace*{-0.1cm}
\begin{eqnarray}\label{eq:53}
\Delta\bar{\tilde{\mathbf G}}_{0,0}&=&\bar {\mathbf W}_0^H\tilde {\mathbf W}_0+\bar {\mathbf W}_1^H\tilde {\mathbf W}_1+\bar {\mathbf W}_2^H\tilde {\mathbf W}_2 \nonumber \\
\Delta\bar{\tilde{\mathbf G}}_{0,1}&=&\bar {\mathbf W}_1^H\tilde {\mathbf W}_0+\bar {\mathbf W}_2^H\tilde {\mathbf W}_1 \nonumber \\
\Delta\bar{\tilde{\mathbf G}}_{0,2}&=&\bar {\mathbf W}_2^H\tilde {\mathbf W}_0 \nonumber \\
\Delta\bar{\tilde{\mathbf G}}_{0,j}&=& \mathbf 0 \quad\quad\textrm{for}\quad 3\le j \le M-1
\end{eqnarray}
\subsection{\textit{Compensating the Real Branch Signal:}} The real branch signal is affected by ISI and ICI terms as shown in (\ref{eq:47}). The proposed algorithm can compensate these two interferences as follows
\subsubsection{\textit{Compensating the ISI}} The third and fourth terms in (\ref{eq:47}) are the ISI terms caused by the $\mathcal{I}$ and $\mathcal{Q}$ branch symbols. Using (\ref{eq:52}) and (\ref{eq:53}), we can compensate these ISI terms at the receiver side as
\begin{eqnarray}\label{eq:54}
\mathbf{\bar u}_{0,comp}\!\!\!\!\!&=&\!\!\!\!\!\Re\{\mathbf  {\bar u}_0\}\!-\!\!\!\sum_{i=1}^{M-1}\!\!\!\Re\{\boldsymbol{\bar\Phi}^H_{k,0}\mathcal{F}_k{\Delta\bar{\bar{\mathbf G}}_{k,0,i} \mathcal{F}_k^H \!\mathbf{\bar\Phi}_{k,i}}\!\}\mathbf {\bar s}_i\nonumber \\
&&-\sum_{i=0}^{M-1}\Re\{\boldsymbol{\bar\Phi}^H_{k,0}\mathcal{F}_k{\Delta\bar{\tilde{\mathbf G}}_{k,0,i} \mathcal{F}_k^H \mathbf{ \tilde\Phi}_{k,i}}\}\mathbf {\tilde s}_i \nonumber \\
\!\!\!\!\!&=&\!\!\!\!\! \mathbf {\bar s}_0+\Re\{\boldsymbol{\bar\Phi}^H_{k,0}\mathcal{F}_k{\Delta\bar{\bar{\mathbf G}}_{k,0,0} \mathcal{F}_k^H \mathbf{ \bar\Phi}_{k,0}}\}\mathbf {\bar s}_0 \nonumber \\
\!\!\!\!\!&=&\!\!\!\!\! [\mathbf I+\Re\{\boldsymbol{\bar\Phi}^H_{k,0}\mathcal{F}_k{\Delta\bar{\bar{\mathbf G}}_{k,0,0} \mathcal{F}_k^H \mathbf{ \bar\Phi}_{k,0}}\}]\mathbf {\bar s}_0 \nonumber \\
\!\!\!\!\!&=&\!\!\!\!\! \Re[\mathbf I+\boldsymbol{\bar\Phi}^H_{k,0}\mathcal{F}_k{\Delta\bar{\bar{\mathbf G}}_{k,0,0} \mathcal{F}_k^H \mathbf{ \bar\Phi}_{k,0}}]\mathbf {\bar s}_0
\end{eqnarray}
\begin{figure*}[t]
	\centering
	\begin{subfigure}[b]{0.275\textwidth}
		\includegraphics[width=\textwidth]{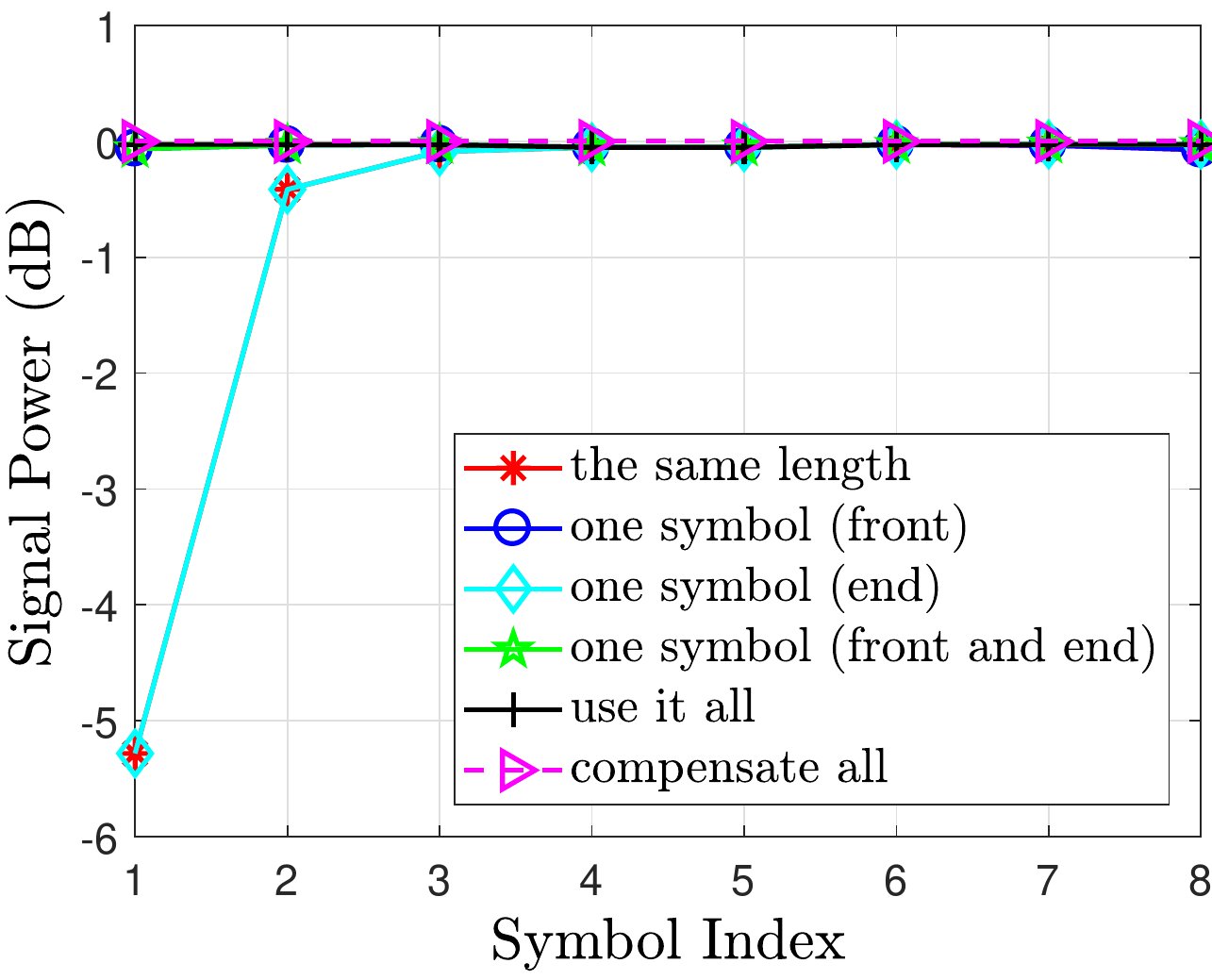}
		\caption{Signal power in $\mathcal{I}$ branch}\label{4a}
	\end{subfigure}
	~ 
	\begin{subfigure}[b]{0.275\textwidth}
		\includegraphics[width=\textwidth]{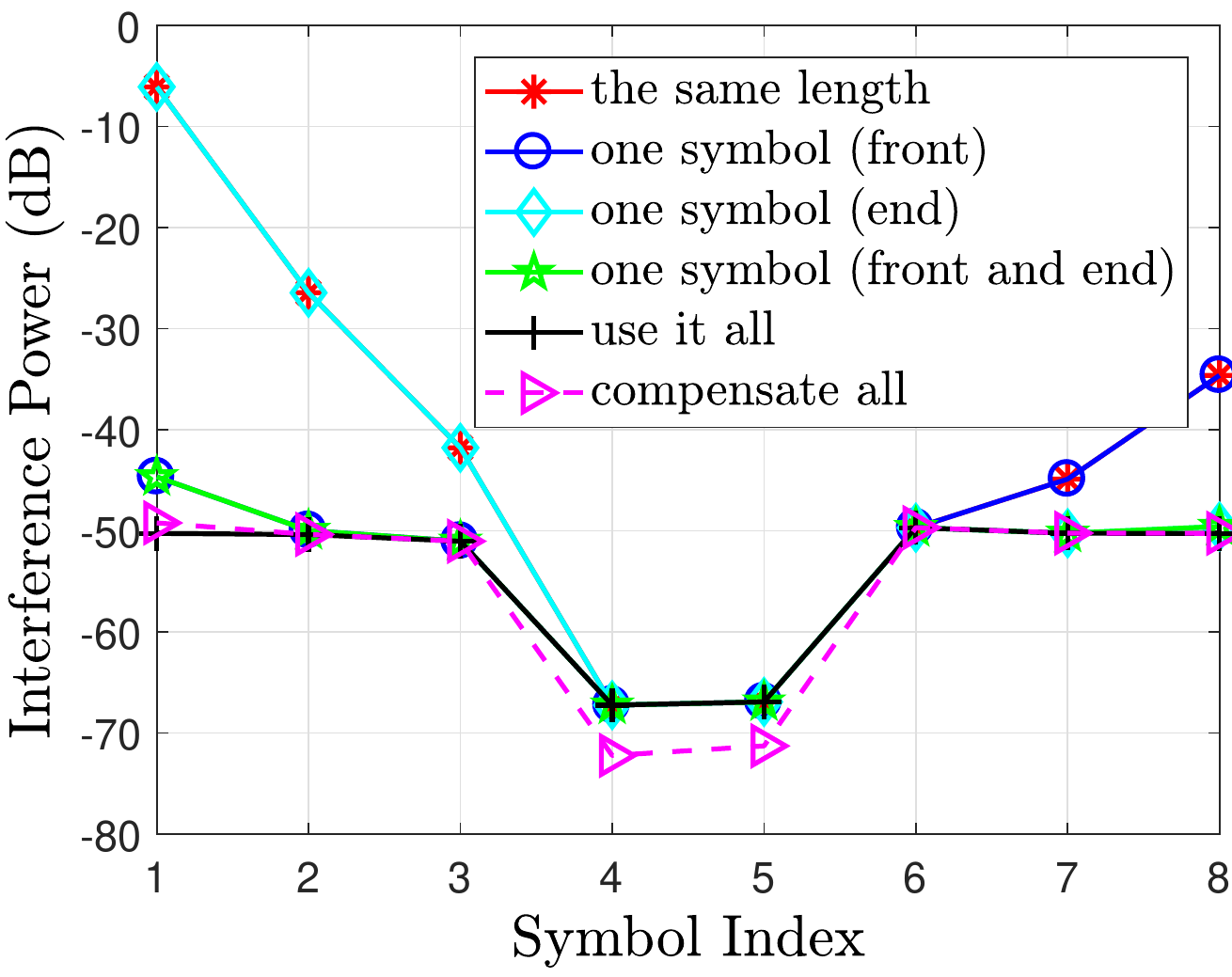}
		\caption{Interference power in $\mathcal{I}$ branch}\label{4b}
	\end{subfigure}
	~ 
	\begin{subfigure}[b]{0.275\textwidth}
		\includegraphics[width=\textwidth]{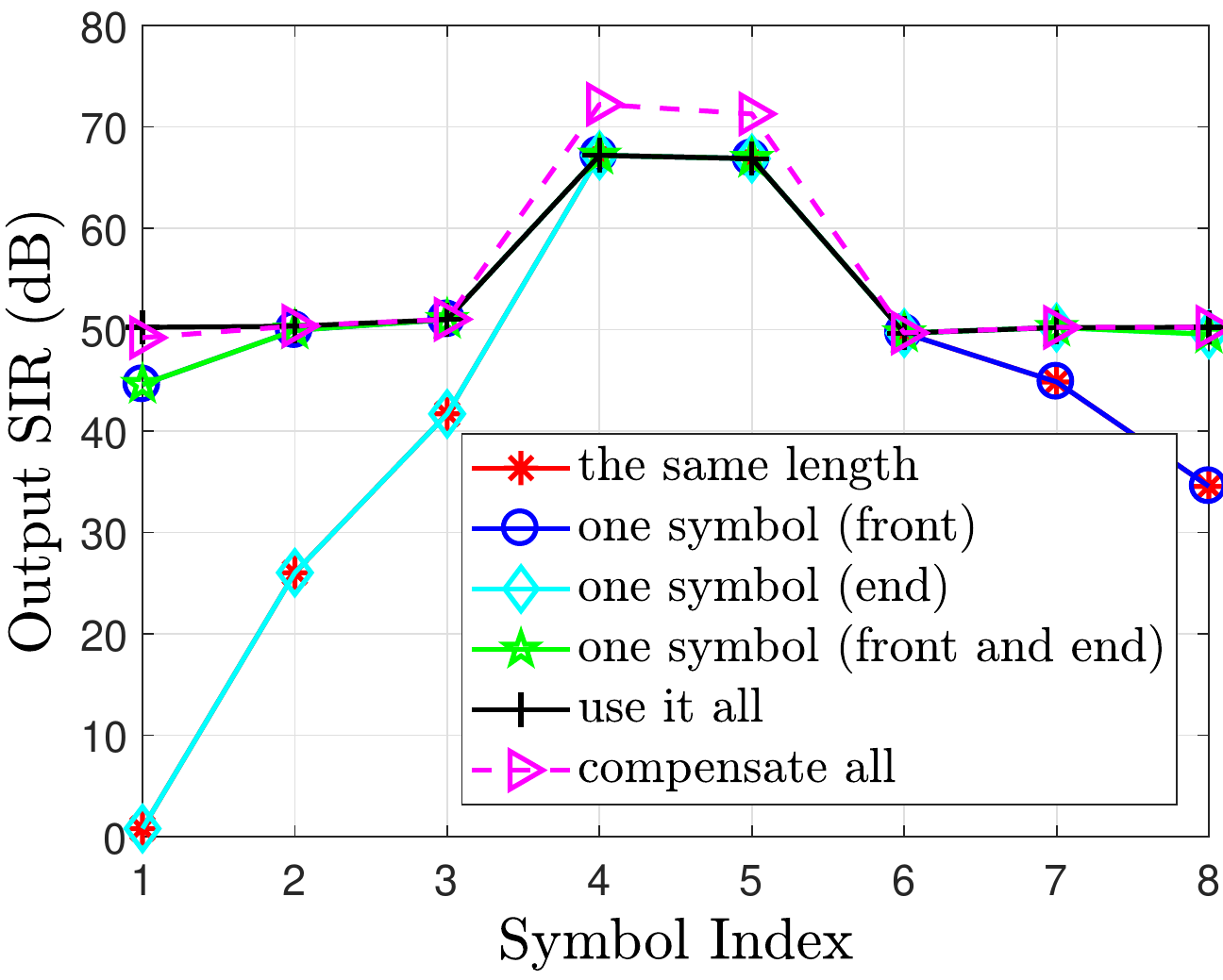}
		\caption{Output SIR in $\mathcal{I}$ branch}\label{4c}
	\end{subfigure}
	~ 
	
	\begin{subfigure}[b]{0.275\textwidth}
		\includegraphics[width=\textwidth]{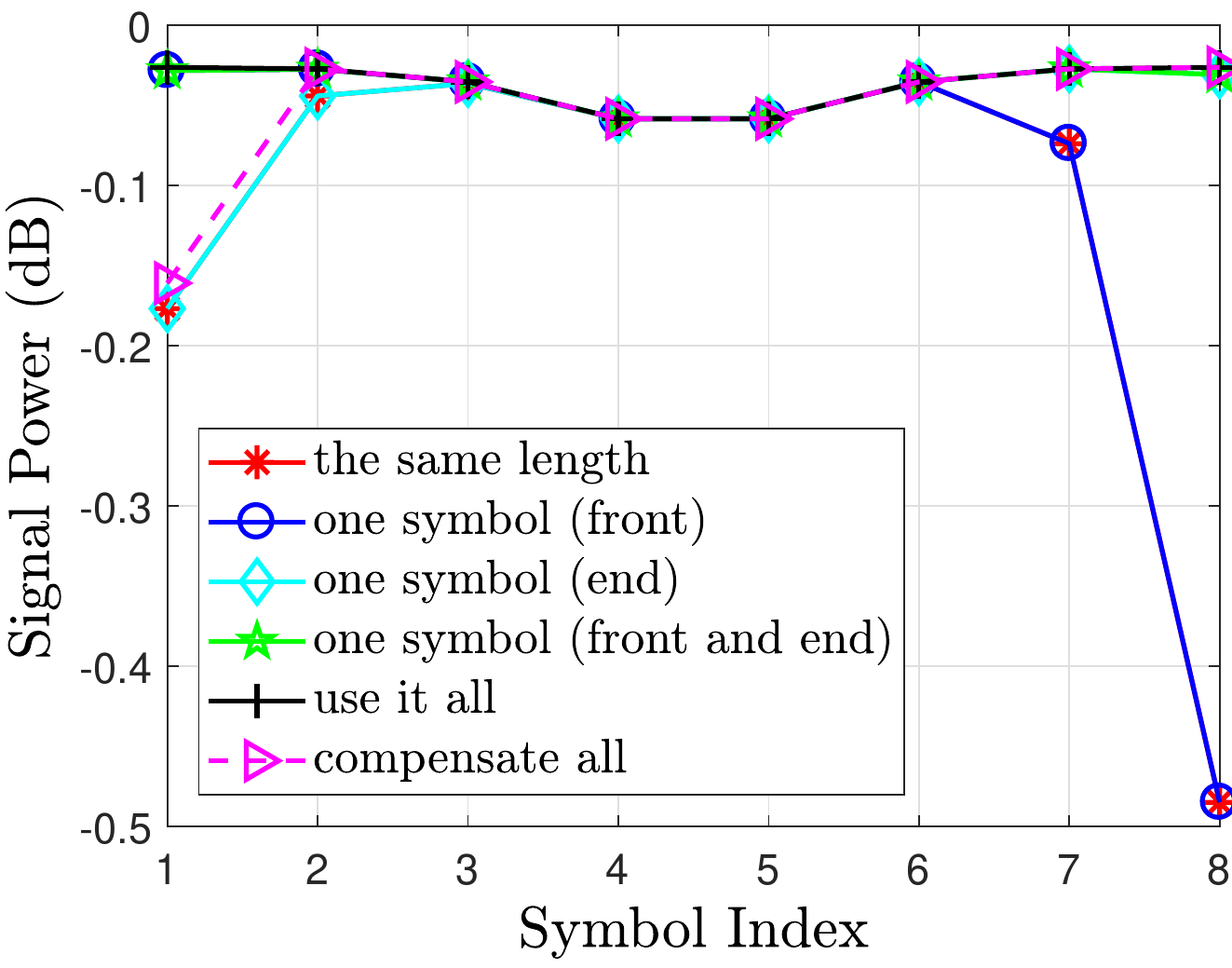}
		\caption{Signal power in $\mathcal{Q}$ branch}\label{4d}
	\end{subfigure}
	~ 
	\begin{subfigure}[b]{0.275\textwidth}
		\includegraphics[width=\textwidth]{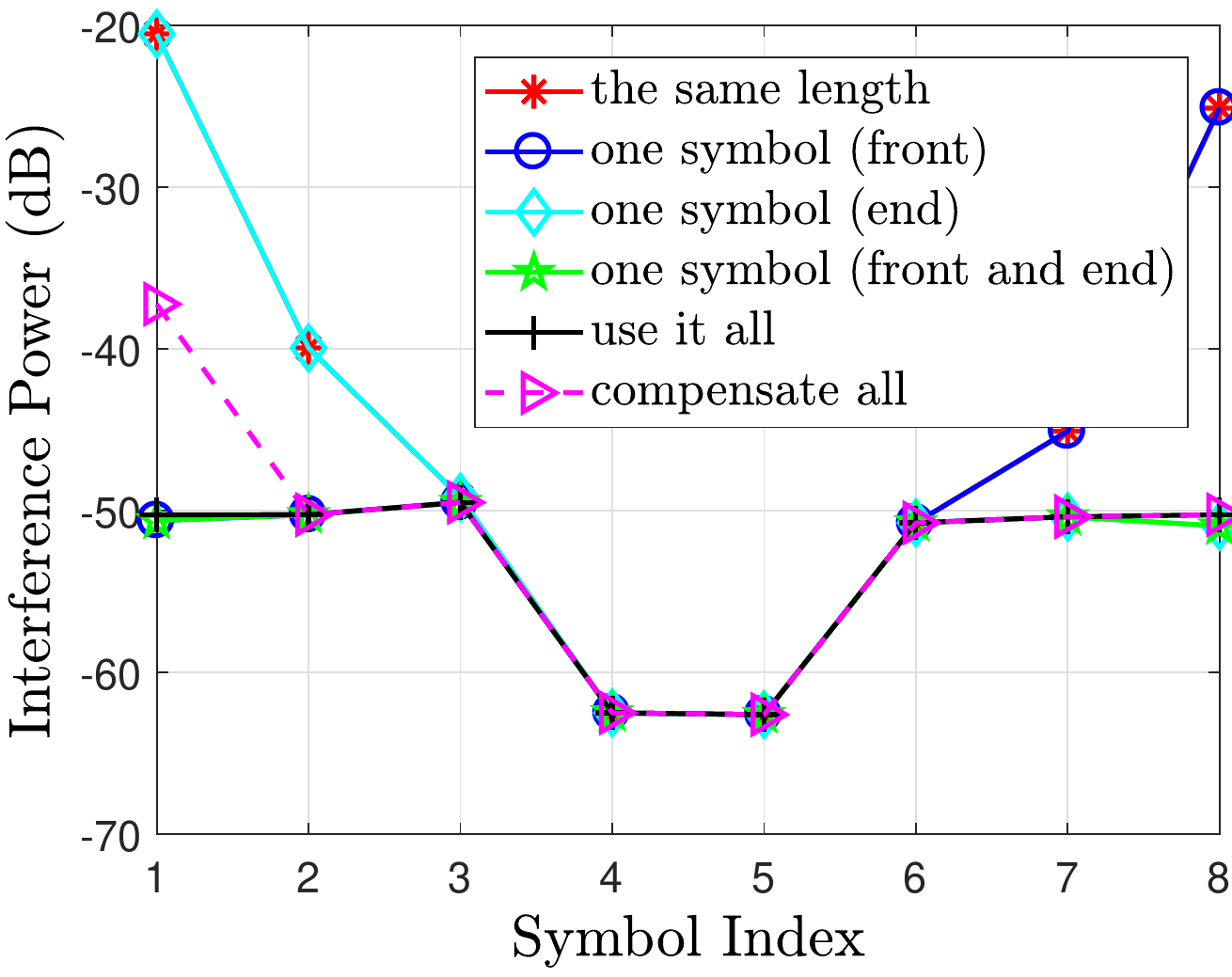}
		\caption{Interference power in $\mathcal{Q}$ branch}\label{4e}
	\end{subfigure}
	~ 
	\begin{subfigure}[b]{0.275\textwidth}
		\includegraphics[width=\textwidth]{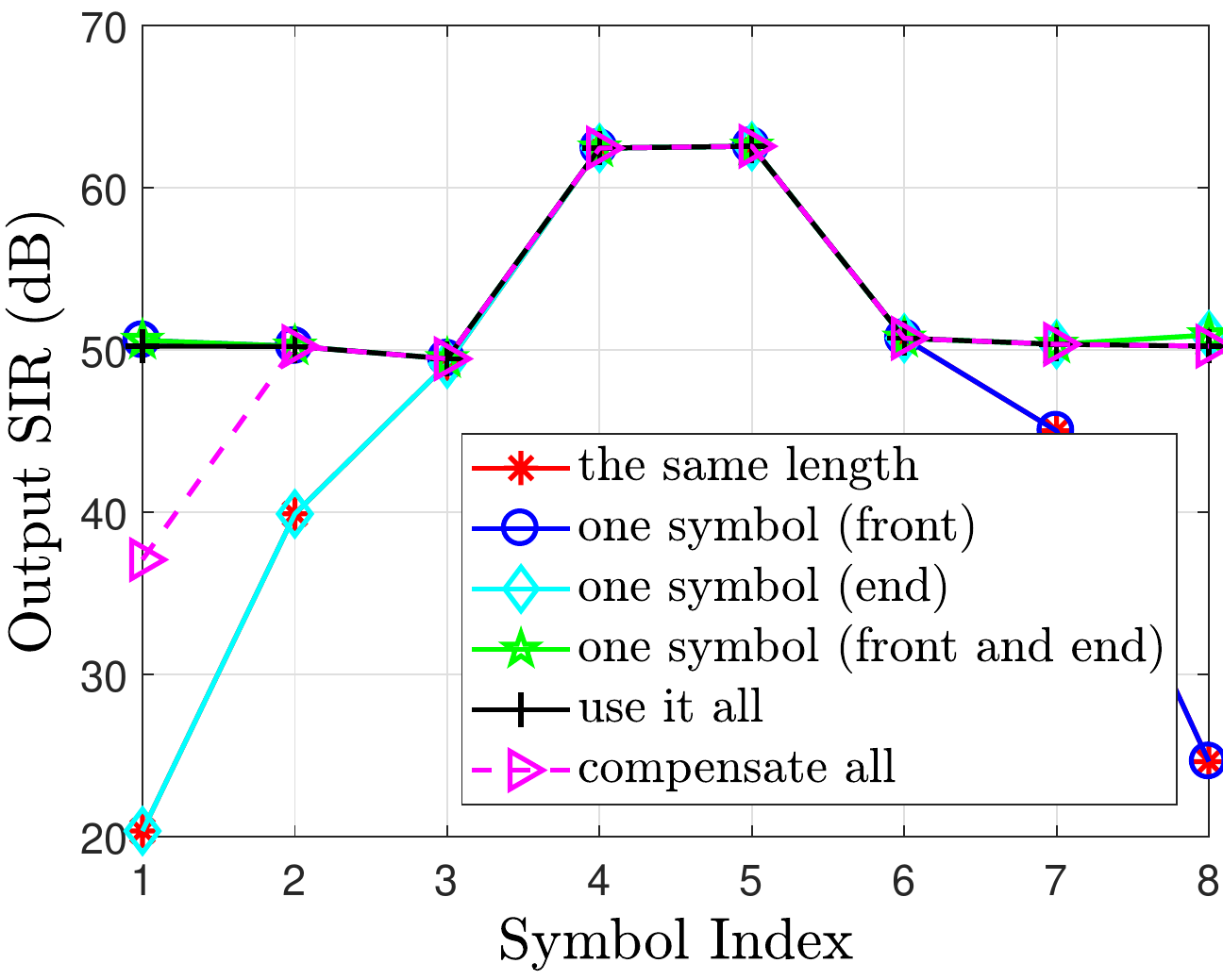}
		\caption{Output SIR in $\mathcal{Q}$ branch}\label{4f}
	\end{subfigure}
	\caption{{Signal and interference power with output SIR using compensation algorithm $(\!K\!=\!6\!)$}}\label{fig:fig6}
\end{figure*}
\subsubsection{\textit{Compensating the ICI}} Apparently, the term $\boldsymbol{\bar\Phi}^H_{k,0}\mathcal{F}_k{\Delta\bar{\bar{\mathbf G}}_{k,0,0} \mathcal{F}_k^H \mathbf{\bar\Phi}_{k,0}}$ in (\ref{eq:54}) is also known. Hence, we can compensate this term by using a ZF (or if we consider the noise term in (\ref{eq:47}) we can use MMSE) equalization at the receiver to estimate $\mathbf {\bar s}_0$ with relatively higher SIR as
\begin{eqnarray}\label{eq:55}
\mathbf {\hat{\bar s}}_{0}\!=\!(\!\Re[\mathbf I\!+\!\boldsymbol{\bar\Phi}^H_{k,0}\mathcal{F}_k{\Delta\bar{\bar{\mathbf G}}_{k,0,0} \mathcal{F}_k^H \!\mathbf{\bar\Phi}_{k,0}}])^{-1}\mathbf {\bar u}_{0,comp}
\end{eqnarray}
\indent It can be seen from (\ref{eq:55}) that both ICI and ISI can be compensated at the receiver side. The proposed compensation algorithm can provide the same SIR for the first real symbol as in the \textit{use it all} case by compensating the effect of FOT as can be seen from Fig. \ref{4c}. Further, we can derive a generalized expression of (\ref{eq:55}) which can be used to further improve the SIR of other symbols as well by compensating their ICI and ISI terms. The generalized expression of (\ref{eq:54}) can be derived as
\begin{eqnarray}\label{eq:56}
\mathbf{\bar u}_{m,comp}\!\!\!\!\!\!&=&\!\!\!\!\!\!\Re\{\!\mathbf  {\bar u}_m\!\}\!-\!\!\!\!\!\!\!\sum_{i=0, i\ne m}^{M-1}\!\!\!\!\!\!\Re\{\boldsymbol{\bar\Phi}^H_{k,m}\mathcal{F}_k{\Delta\!\bar{\bar{\mathbf G}}_{k,m,i} \mathcal{F}_k^H \!\mathbf{\bar\Phi}_{k,i}}\!\}\mathbf {\bar s}_i \nonumber \\ 
&&-\sum_{i=0}^{M-1}\!\!\Re\{\boldsymbol{\bar\Phi}^H_{k,m}\mathcal{F}_k{\Delta\bar{\tilde{\mathbf G}}_{k,m,i} \mathcal{F}_k^H \mathbf{ \tilde\Phi}_{k,i}}\}\mathbf {\tilde s}_i \nonumber \\
\!\!\!\!\!&=&\!\!\!\!\! \mathbf {\bar s}_m\!+\!\Re\{\boldsymbol{\bar\Phi}^H_{k,m}\mathcal{F}_k{\Delta\bar{\bar{\mathbf G}}_{k,m,m} \mathcal{F}_k^H \mathbf{ \bar\Phi}_{k,m}}\}\mathbf {\bar s}_m \nonumber \\
\!\!\!\!\!&=&\!\!\!\!\! [\mathbf I\!+\!\Re\{\boldsymbol{\bar\Phi}^H_{k,m}\mathcal{F}_k{\Delta\bar{\bar{\mathbf G}}_{k,m,m} \mathcal{F}_k^H \mathbf{ \bar\Phi}_{k,m}}\}]\mathbf {\bar s}_m \nonumber \\
\!\!\!\!\!&=&\!\!\!\!\! \Re[\mathbf I\!+\!\boldsymbol{\bar\Phi}^H_{k,m}\mathcal{F}_k{\Delta\bar{\bar{\mathbf G}}_{k,m,m} \mathcal{F}_k^H \mathbf{ \bar\Phi}_{k,m}}]\mathbf {\bar s}_m
\end{eqnarray}
Similarly, (\ref{eq:55}) can be generalized as
\begin{eqnarray}\label{eq:57}
\mathbf {\hat{\bar s}}_{m}\!\!=\!(\!\Re[\mathbf I\!\!+\!\!\boldsymbol{\bar\Phi}^H_{k,m}\mathcal{F}_k\!{\Delta\!\bar{\bar{\mathbf G}}_{k,m,m}\! \mathcal{F}_k^H\! \mathbf{\bar \Phi}_{k,m}}])^{\!-\!1}\!\mathbf {\bar u}_{m,comp}
\end{eqnarray}
In (\ref{eq:56}), it is worth mentioning that the term $\sum_{i=0, i\ne m}^{M-1}\Re\{\boldsymbol{\bar\Phi}^H_{k,m}\mathcal{F}_k{\Delta\bar{\bar{\mathbf G}}_{k,m,i} \mathcal{F}_k^H \mathbf{\bar\Phi}_{k,i}}\}\mathbf {\bar s}_i$ should be treated carefully for $i=0$, since only accurate $\mathbf {\hat{\bar s}}_{0}$ will bring accurate compensation to other symbols, otherwise errors will be introduced, which implies that $\mathbf {\hat{\bar s}}_{0}$ should be always compensated first.
\subsection{\textit{Compensating the Imaginary Branch Signal:}} The equalized imaginary branch symbol, $\mathbf  {\tilde u}_m$ can be written as follows using the same approach as adopted for the real branch.
\begin{eqnarray}\label{eq:58}
\mathbf  {\tilde u}_m \!\!\!\!&=&\!\!\!\! \boldsymbol{\tilde\Phi}^H_{k,m}\mathcal{F}_k{\tilde{\tilde{\mathbf G}}_{k,m,m} \mathcal{F}_k^H \mathbf{ \tilde\Phi}_{k,m}}\mathbf {\tilde s}_m\nonumber \\
&&+\sum_{i=0, i\neq m}^{M-1}\underbrace{\boldsymbol{\tilde\Phi}^H_{k,m}\mathcal{F}_k{\tilde{\tilde{\mathbf G}}_{k,m,i} \mathcal{F}_k^H \mathbf{\tilde \Phi}_{k,i}}}_{\tilde{\tilde{\mathbf Q}}_{k,m,i}}\mathbf {\tilde s}_i \nonumber \\
&&+\sum_{i=0}^{M-1}\underbrace{\boldsymbol{\tilde\Phi}^H_{k,m}\mathcal{F}_k{\tilde{\bar{\mathbf G}}_{k,m,i} \mathcal{F}_k^H \mathbf{ \bar\Phi}_{k,i}}}_{\tilde{\bar{\mathbf Q}}_{k,m,i}}\mathbf {\bar s}_i\nonumber \\
&&+ \underbrace{\mathbf E_m\boldsymbol{\tilde\Phi}^H_{k,m}\mathbf {\mathcal{F}}_k\mathbf{\tilde{P}}^{H}_{k,m}\mathbf{n}}_{\mathbf{\tilde u}_{noise,m}}
\end{eqnarray}
According to the orthogonality of FBMC with infinite filter length \cite{7549072}, $\tilde{\tilde{\mathbf Q}}_{k,m,i}$ and $\tilde{\bar{\mathbf Q}}_{k,m,i}$ have the following property
\begin{eqnarray}\label{eq:59}
\tilde{\tilde{\mathbf Q}}_{k,m,i} \!\!\!\!\!&=&\!\!\!\!\! \left\{\!\!
            \begin{array}{lcl}
             j\mathbf {I} + \Re\{{\tilde{\tilde{\mathbf Q}}_{k,m,i}}\} \!\!\! \quad \textrm{for} \quad  i=m\\
             \Re\{{\tilde{\tilde{\mathbf Q}}_{k,m,i}}\}  \!\!\quad \quad \quad \textrm{for} \quad i \neq m
             \end{array}
        \right.\nonumber \\
\tilde{\bar{\mathbf Q}}_{k,m,i}\!\!\!\!\! &=& \!\!\!\!\! \Re\{{\tilde{\bar{\mathbf Q}}_{k,m,i}}\} \!\!\!\!\!\! \quad \quad \textrm{for} \quad i = 0,\cdots,M\!-\!1
\end{eqnarray}
Using the property of infinite filter length given in (\ref{eq:59}), we can write (\ref{eq:58}) as
\begin{eqnarray}\label{eq:60}
\mathbf  {\tilde u}_m \!\!\!\!\!&=& \!\!\!\!\!j\mathbf {\tilde s}_m\!\!+\!\!\underbrace{\Big[\!\!\sum_{i=0}^{M-1}\!\!\Re\{{\tilde{\tilde{\mathbf Q}}_{k,m,i}}\}\mathbf {\tilde s}_i\!+\!\!\!\!\sum_{i=0}^{M-1}\!\!\Re\{{\tilde{\bar{\mathbf Q}}_{k,m,i}}\}\mathbf {\bar s}_i\Big]}_{\mathbf {\tilde u}_{intri,m}}\nonumber \\
&&+ \mathbf{\tilde u}_{noise,m}
\end{eqnarray}
Taking the imaginary part of (\ref{eq:60}), we have
\begin{eqnarray}\label{eq:61}
\Im\{\mathbf  {\tilde u}_m\} \!\!\!\!\!&=&\!\!\!\!\! \mathbf {\tilde s}_m\!+\!\Im\{\mathbf{\tilde u}_{noise,m}\}
\end{eqnarray}
\indent We can now compensate the $\mathcal{Q}$ branch as well using the same approach used for the $\mathcal{I}$ branch; however, since all the symbols already have good initial SIR, it will be easier to compensate them in this branch compared to the $\mathcal{I}$ branch. The compensation approach for the $\mathcal{Q}$ branch is as follows
\begin{eqnarray}\label{eq:62}
\mathbf {\hat{\tilde s}}_{m}\!=\!(\!\Im[\mathbf I\!+\!\boldsymbol{\tilde\Phi}^H_{k,m}\mathcal{F}_k{\Delta\!\tilde{\tilde{\mathbf G}}_{k,m,m} \mathcal{F}_k^H \!\mathbf{\tilde \Phi}_{k,m}}])^{\!-\!1}\!\mathbf {\tilde u}_{m,comp}
\end{eqnarray}
where,
\begin{eqnarray}\label{eq:63}
\mathbf{\tilde u}_{m,comp}\!\!\!\!\!&=&\!\!\!\!\!\Im\{\mathbf  {\tilde u}_m\}\!-\!\!\!\!\!\!\sum_{i=0, i\ne m}^{M-1}\!\!\!\!\!\!\Im\{\!\boldsymbol{\tilde\Phi}^H_{k,m}\mathcal{F}_k{\Delta\!\tilde{\tilde{\mathbf G}}_{k,m,i} \mathcal{F}_k^H \!\mathbf{\tilde\Phi}_{k,i}}\}\mathbf {\tilde s}_i \nonumber \\
&&-\sum_{i=0}^{M-1}\!\!\!\Im\{\boldsymbol{\tilde\Phi}^H_{k,m}\mathcal{F}_k{\Delta\!\tilde{\bar{\mathbf G}}_{k,m,i} \mathcal{F}_k^H \!\mathbf{ \bar\Phi}_{k,i}}\}\mathbf {\bar s}_i
\end{eqnarray}
\subsection{\textit{Combining Real and Imaginary Branches:}}
With (\ref{eq:57}) and (\ref{eq:62}), we can write the compensated estimation of $\mathbf {s}_{m}$ as follows
\begin{eqnarray}\label{eq:64}
\mathbf  {\hat s}_{m}\!\!\!\!\!&=&\!\!\!\!\!\mathbf {\hat{\bar s}}_{m} \!+\! j \mathbf {\hat{\tilde s}}_{m} 
\end{eqnarray}
\section{Simulation Results}\label{sec:simulation}
\begin{figure}[h]
	\begin{center}
		\includegraphics[scale=0.65]{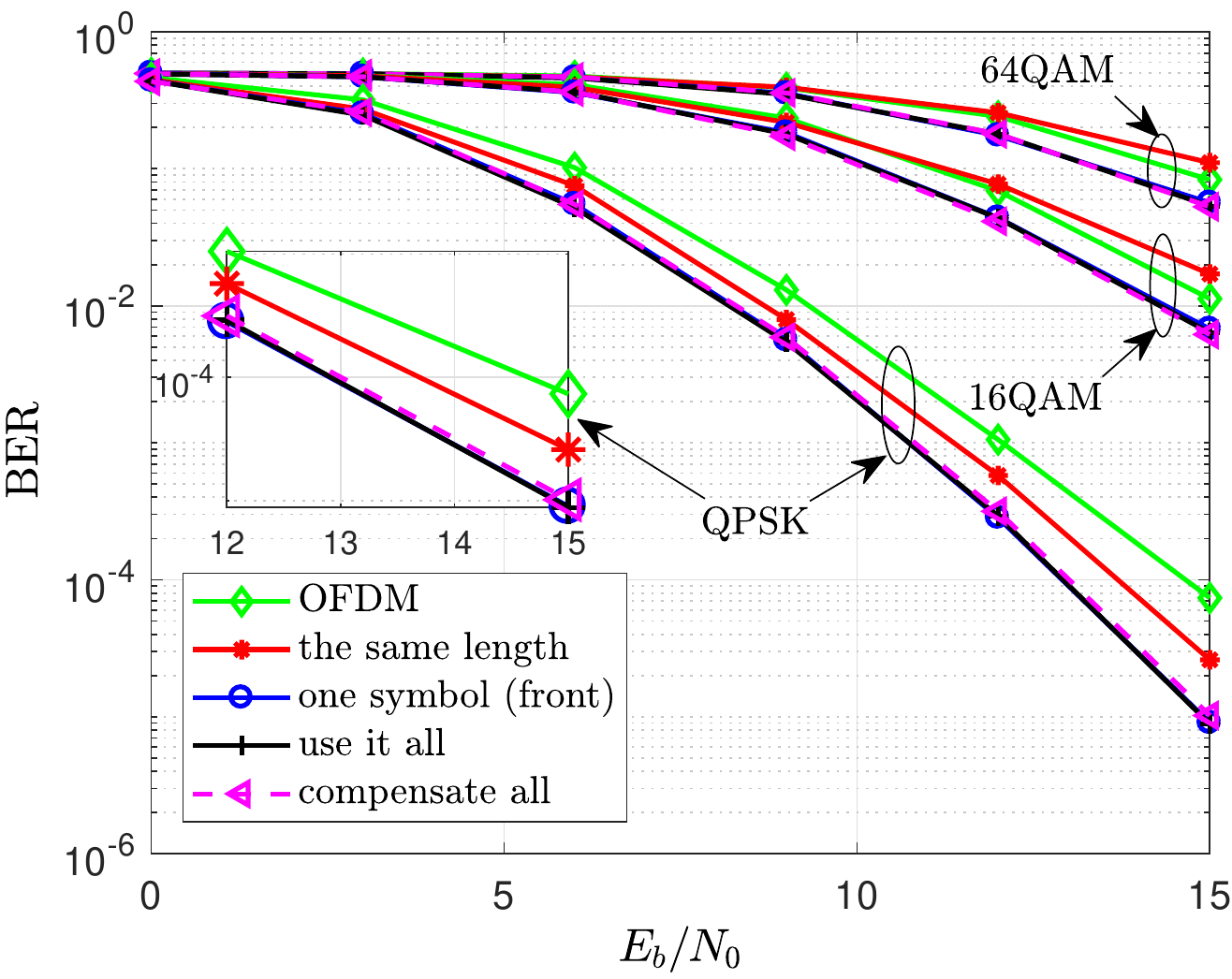}
		\caption{{BER performance of OFDM and FBMC system with and without compensation}}\label{fig:fig8}
	\end{center}
\end{figure}

In this section, we present a set of simulation results to demonstrate the effectiveness of the proposed compensation algorithm in the \textit{the same length} case. For simulations, the selected parameters for the MIMO-FBMC system includes the IOTA prototype filter with overlapping factor $K\!=\!6$. The number of transmit and receive antennas are $N_t\!=\!N_r\!=\!2$. The desired signal is modulated by QPSK with normalized power and input signal to noise ratio (SNR) is controlled by the noise power. The LTE channel model considered in our simulation is the extended pedestrian A model (EPA) \cite{dahlman20134g}. For the equalization, the MMSE based equalizer is selected as it is more generic. As we have concluded in Section \ref{sec5} that our main concern is the first real branch symbol which has a very low SIR value of around 2dB. The proposed compensation algorithm given in (\ref{eq:55}) significantly improves the SIR of the first real symbol i.e. the signal power increase from -5.1dB to 0dB while the interference level drops from -5dB to -48dB. Hence, increasing the SIR of the first real symbol from 2dB to 48dB as can be seen from Fig. {\ref{fig:fig6}}. Note that we do not need to compensate the imaginary branch as it already has sufficient SIR values for detecting all the symbols at the receiver as discussed in Section \ref{sec5}.\vspace*{.1cm}\\
\indent However, the term \textit{acceptable SIR value} is strongly dependent on the modulation order as higher modulation order require high SIR values for achieving a specific required performance. The proposed general form of the compensation algorithm given in (\ref{eq:57}) and (\ref{eq:62}) is incorporated at the receiver side of the MIMO-FBMC system. The algorithm significantly improves the SIR of all the symbols in the real and imaginary branches respectively as can be see from Fig. {\ref{fig:fig6}}. Compensating all the symbols can help in improving the probability of detection at the receiver. The coded results (convolutional code with code rate 1/2) for the BER performance of various FOT schemes in MIMO-FBMC system with and without compensation algorithm are presented in Fig. {\ref{fig:fig8}}. It can be observed that the system performance in case of \textit{use it all} and \textit{one symbol (front)} has similar BER performance but the latter required only one extra tail compared to the former, which required ${K\!-\!1}$ extra tails. The \textit{same length} case requires no extra tail but has a relatively poor BER performance.
\\
\indent We have used conventional MIMO-OFDM as a baseline scheme to show the advantage of the proposed algorithm over such conventional multicarrier schemes. For a fair comparison between MIMO-FBMC and MIMO-OFDM systems, the SNR loss, due to the cyclic prefix (overhead) in OFDM, must be considered. For this reason, we have calculated the noise power for both systems as discussed in \cite{7390479}. The comparison shows the significance of the proposed algorithm especially for higher modulation schemes. It can be seen from Fig. {\ref{fig:fig8}} that for low order modulation schemes like QPSK, MIMO-FBMC system without compensation can still perform better than conventional MIMO-OFDM but if we increase the modulation schemes to higher order like 16QAM or 64QAM, the performance of MIMO-FBMC system without compensation becomes poorer than MIMO-OFDM due to self-interference caused by FOT. In such cases, use of the proposed compensation algorithm is very significant as it not only provides better performance but also improves the spectral efficiency (SE) of the system.\vspace*{.1cm}
\begin{figure}[h]
	\centering
	\begin{subfigure}[b]{0.24\textwidth}
		\includegraphics[width=\textwidth]{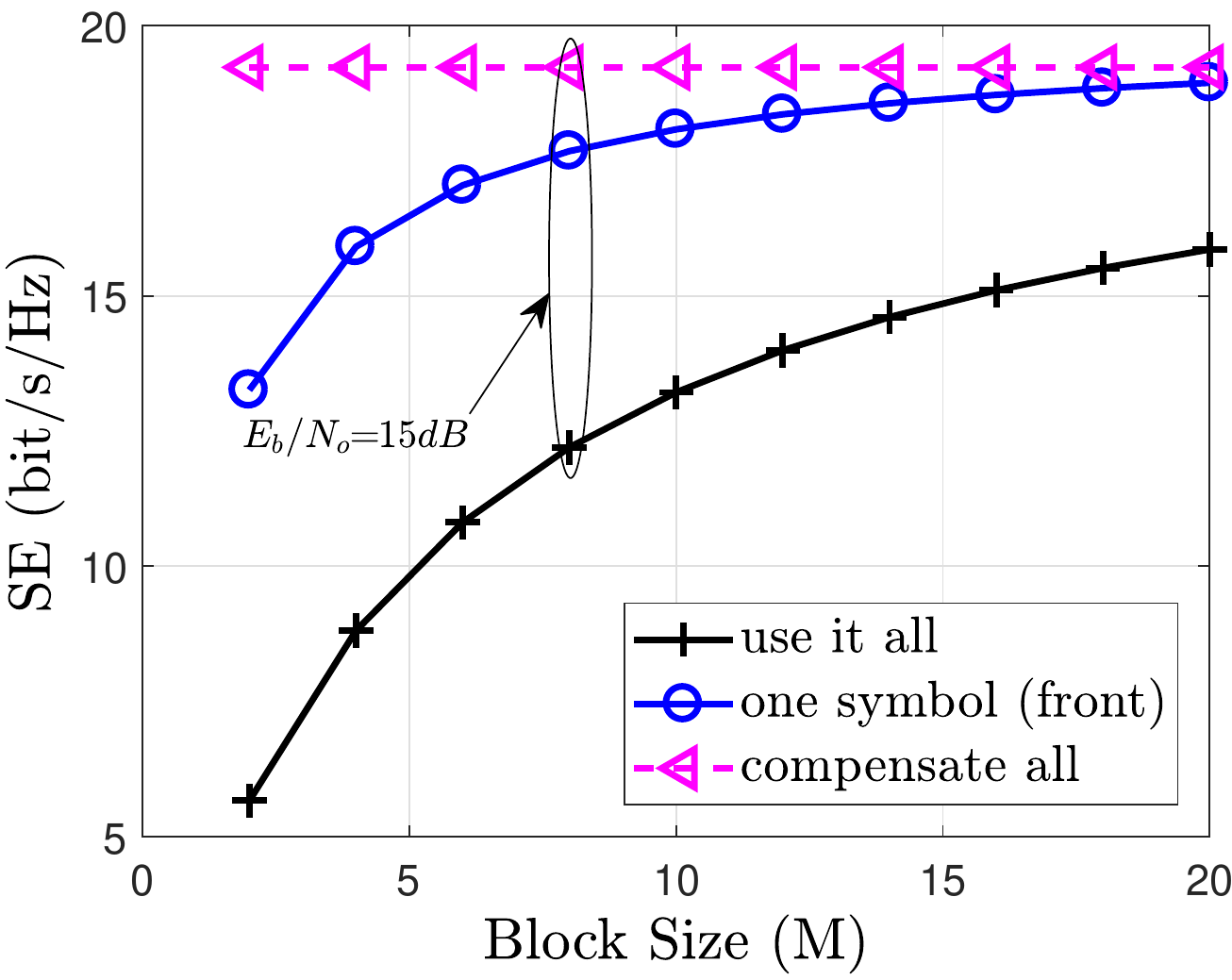}
		\caption{{SE with respect to $M$}}
		\label{6a}
	\end{subfigure}\hspace{-0.15\baselineskip}
	\begin{subfigure}[b]{0.24\textwidth}
		\includegraphics[width=\textwidth]{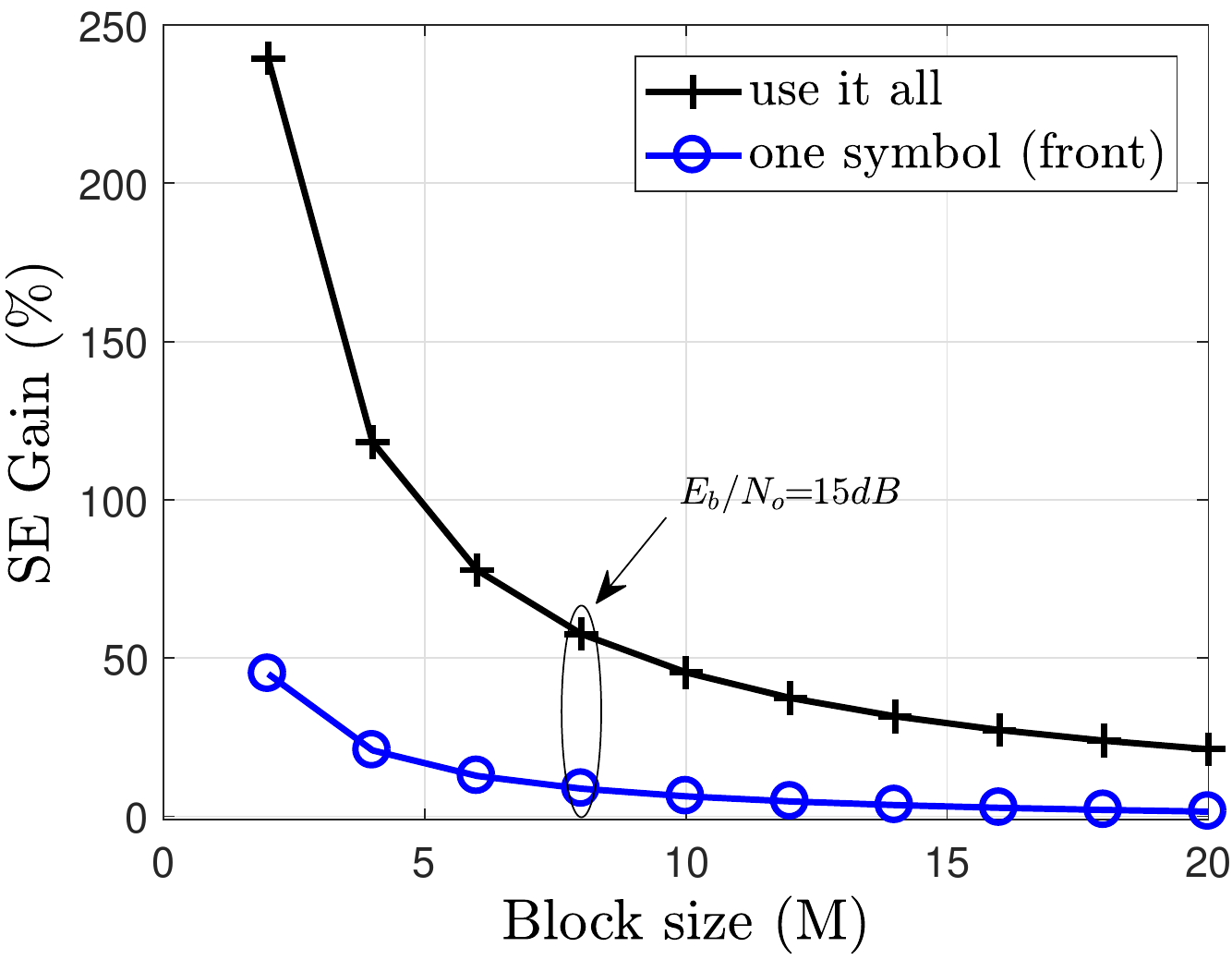}
		\caption{{SE gain with respect to $M$}}
		\label{6b}
	\end{subfigure}
	\caption{{Spectral efficiency of MIMO-FBMC with respect to block size ($M$)}}\label{fig:fig9}
\end{figure}
\begin{figure}[h]
	\centering
	\begin{subfigure}[b]{0.24\textwidth}
		\includegraphics[width=\textwidth]{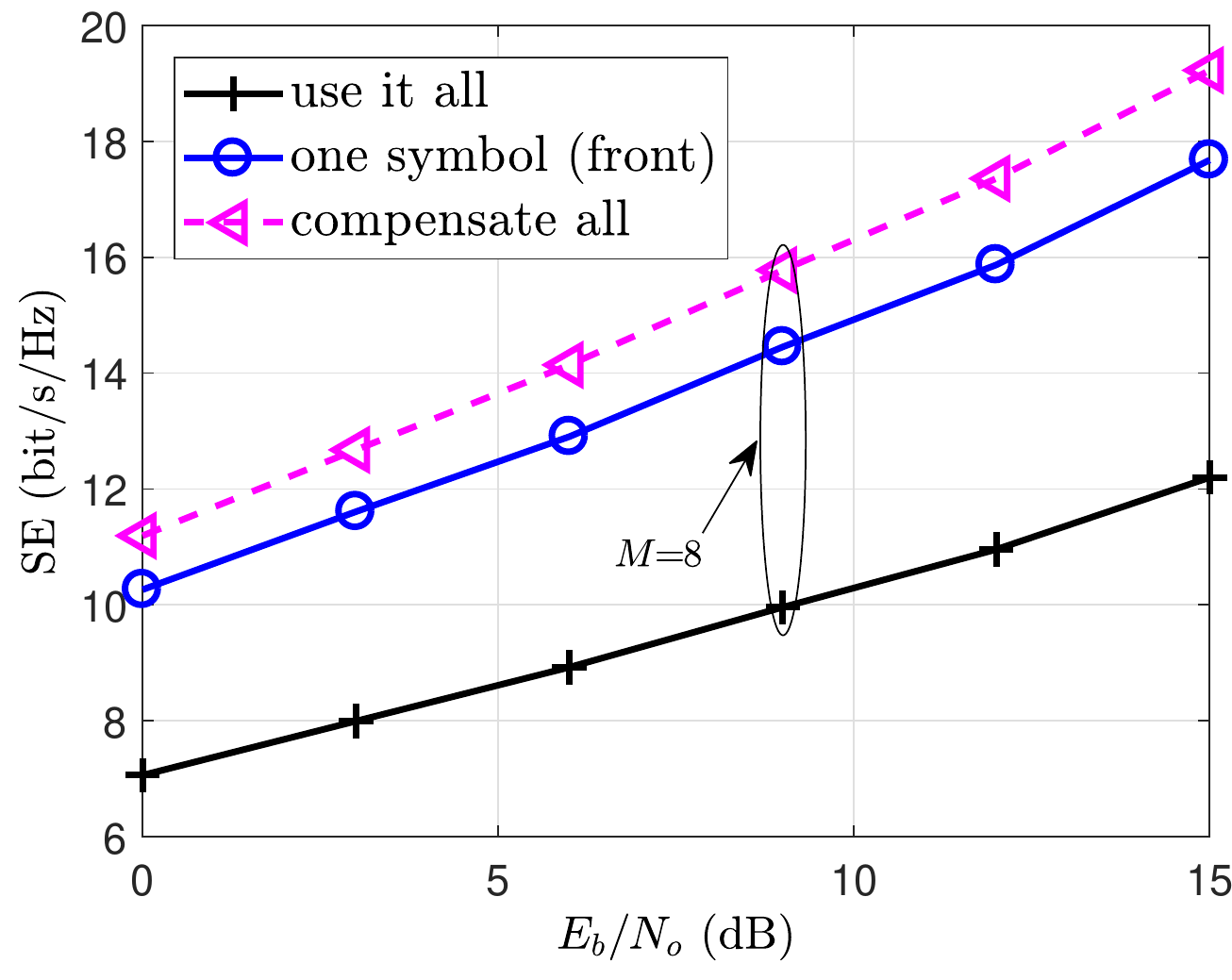}
		\caption{{SE with respect to $E_b/N_o$}}
		\label{6c}
	\end{subfigure}\hspace{-0.15\baselineskip}
	\begin{subfigure}[b]{0.24\textwidth}
		\includegraphics[width=\textwidth]{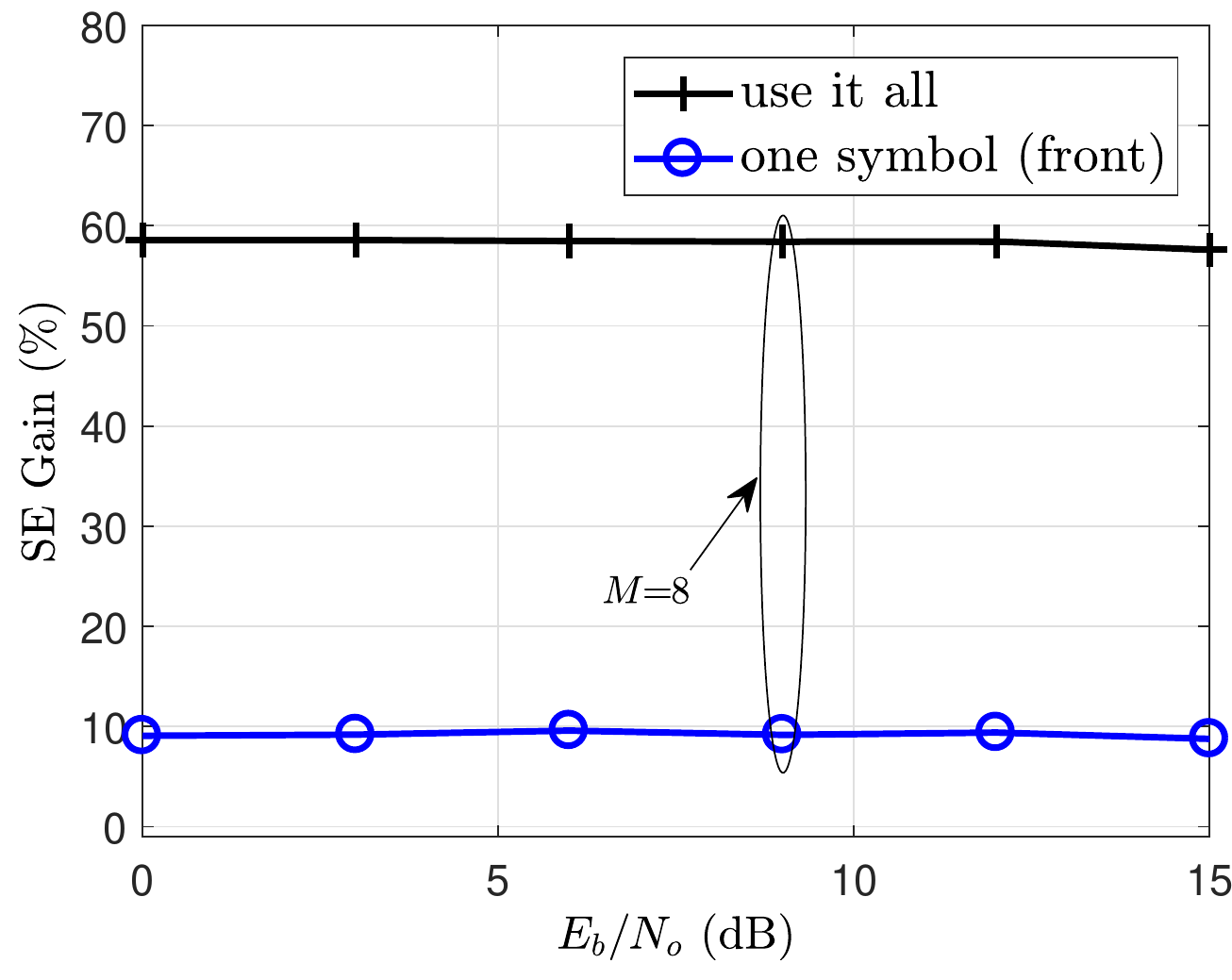}
		\caption{{SE gain with respect to $E_b/N_o$}}
		\label{6d}
	\end{subfigure}
	\caption{{Spectral efficiency of MIMO-FBMC with respect to SNR ($E_b/N_o$)}}\label{fig:fig10}
\end{figure}
\\\indent The SE of the system has been simulated using Shannon equation \cite{tse2005fundamentals} which gives an upper bound of the capacity that the system can achieve i.e. maximum error free transmission rate.
\textit{Note that the capacity is measured using only the simulation and is not the exact representation of achievable capacity}. The objective is to provide an idea regarding the spectral efficiency gain that can be achieved using the filter output truncation and the compensation algorithm at the receiver. The spectral efficiency expression with respect to the block size $M$ used in the simulation is given as follows
\begin{eqnarray}
\textrm{SE} = \min \{N_T,N_R\} \!\!\times\!\! \frac{M}{M+\alpha}\! \Bigg\{\!\!\frac{1}{M}\!\sum_{i=1}^{M}\! \log_{2}\!{(1\!\!+\!\!SINR_m\!)}\!\!\Bigg\}
\end{eqnarray}
where $\alpha$ represents the overhead in each case i.e. $K-1$ for \textit{use it all}, 1 for \textit{one symbol (front)} and 0 for \textit{compensate all}.\vspace*{.1cm}\\
\indent The SE in each case is given in Fig. {\ref{fig:fig9}}. It can be observed from Fig. {\ref{6a}} that the SE is independent of $M$ for the \textit{compensate all} case whereas it is dependent for the \textit{use it all} and \textit{one symbol (front)} cases as they have one and $K-1$ tails respectively with each block. It can also be seen from Fig. {\ref{6b}} that the SE gain obtained using the \textit{compensate all} case reduces with the increase in $M$ for both \textit{use it all} and \textit{one symbol (front)} cases. Hence, the compensation algorithm is best suited for applications that has a frame structure based on moderate $M$.
The SE results for a range of SNR ($E_b/N_o$) values are also shown in Fig. {\ref{fig:fig10}}. With a fixed block size, the SE of the system increases as input SNR ($E_b/N_o$) increases as shown in Fig. \ref{6c}. It can be observed that SE performance of the \textit{compensate all} case is better than \textit{one symbol (front)} and \textit{use it all} cases, as these FOT schemes require certain overhead to achieve improved BER performance. However, the proposed compensation algorithm provides similar BER performance by compensating the effects of FOT in the \textit{same length} case without introducing any overhead. This enables \textit{compensate all} case to have a certain SE gain compared to other FOT schemes as can be seen from Fig. \ref{6d}. \vspace*{-.1cm}
\section{Conclusion}
The impact of finite filter length and different types of FOT has been theoretically analyzed in a MIMO-FBMC system. The analysis is based on a compact matrix model of a MIMO-FBMC system, which was then used for investigating the effects of FOT on the detection performance in terms of the SIR and BER. The analysis showed that although FOT can avoid overhead but it also destroys the orthogonality in the FBMC system thus introducing interferences. However, due to the isolation property between the (FBMC) symbols, only real part of the first symbol or the imaginary part of the last symbol are affected by the aforementioned interferences.\vspace*{.1cm} \\
\indent A general form of compensation algorithm based on the observations made in the theoretical analysis has been designed to compensate the symbols in a MIMO-FBMC block to improve the SIR of each symbol. The advantage of this algorithm is that it improves the spectral efficiency of the system as it requires no overhead and at the same time can still achieve similar performance compared to the case without FOT. However, the spectral efficiency gain tends to reduce with the increasing $M$ as the overhead tends to decrease with increase in frame size.\vspace*{.1cm} \\
\indent The proposed analytical framework developed in this paper provide useful insights into the effect of finite filter length and FOT on the system performance and the proposed compensation algorithm can enable MIMO-FBMC system to achieve higher spectral efficiency compared to its conventional counterpart.\vspace*{-.2cm}
\bibliographystyle{IEEEtranCustom}
\bibliography{IEEEabrv,varUD}
\end{document}